\documentclass[]{pasj01}
\Received{2022/05/17}

\usepackage{lineno}
\usepackage{color}
\usepackage[separate-uncertainty, angle-symbol-over-decimal]{siunitx}
\usepackage{url}
\usepackage{bm}
\usepackage[T1]{fontenc}

\begin{document} 

\title{Video Observations of Tiny Near-Earth Objects with Tomo-e Gozen}


\author{Jin \textsc{Beniyama}\altaffilmark{1,2,*}}
\email{beniyama@ioa.s.u-tokyo.ac.jp}
\author{Shigeyuki \textsc{Sako}\altaffilmark{1,3,4}}
\author{Ryou \textsc{Ohsawa}\altaffilmark{1}}
\author{Satoshi \textsc{Takita}\altaffilmark{1}}
\author{Naoto \textsc{Kobayashi}\altaffilmark{1,3,4,5}}
\author{Shin-ichiro \textsc{Okumura}\altaffilmark{6}}
\author{Seitaro \textsc{Urakawa}\altaffilmark{6}}
\author{Makoto \textsc{Yoshikawa}\altaffilmark{7}}
\author{Fumihiko \textsc{Usui}\altaffilmark{7,8}}
\author{Fumi \textsc{Yoshida}\altaffilmark{9,10}}

\author{Mamoru \textsc{Doi}\altaffilmark{1,3,4,11}}
\author{Yuu \textsc{Niino}\altaffilmark{1}}
\author{Toshikazu \textsc{Shigeyama}\altaffilmark{11}}
\author{Masaomi \textsc{Tanaka}\altaffilmark{12,13,14}}
\author{Nozomu \textsc{Tominaga}\altaffilmark{14,15,16}}

\author{Tsutomu \textsc{Aoki}\altaffilmark{5}}
\author{Noriaki \textsc{Arima}\altaffilmark{1,2}}
\author{Ko \textsc{Arimatsu}\altaffilmark{17}}
\author{Toshihiro \textsc{Kasuga}\altaffilmark{15}}
\author{Sohei \textsc{Kondo}\altaffilmark{5}}
\author{Yuki \textsc{Mori}\altaffilmark{5}}
\author{Hidenori \textsc{Takahashi}\altaffilmark{5}}
\author{Jun-ichi \textsc{Watanabe}\altaffilmark{15}}


\altaffiltext{1}{%
Institute of Astronomy, Graduate School of Science,
The University of Tokyo, 2-21-1 Osawa, Mitaka, Tokyo 181-0015, Japan}
\altaffiltext{2}{%
Department of Astronomy, Graduate School of Science,
The University of Tokyo, 7-3-1 Hongo, Bunkyo-ku, Tokyo 113-0033, Japan}
\altaffiltext{3}{%
UTokyo Organization for Planetary Space Science, 
The University of Tokyo, Hongo, Tokyo 113-0033, Japan}
\altaffiltext{4}{%
Collaborative Research Organization for 
Space Science and Technology, 
The University of Tokyo, Hongo, Tokyo 113-0033, Japan}
\altaffiltext{5}{%
Kiso Observatory, 
Institute of Astronomy, Graduate School of Science, 
The University of Tokyo, 10762-30 Mitake, Kiso-machi,
Kiso-gun, Nagano 397-0101, Japan}
\altaffiltext{6}{%
Japan Spaceguard Association, Bisei Spaceguard Center,
1716-3 Okura, Bisei, Ibara, Okayama 714-1411, Japan}
\altaffiltext{7}{%
Institute of Space and Astronautical Science,
Japan Aerospace Exploration Agency, 3-1-1 Yoshinodai,
Chuo-ku, Sagamihara, Kanagawa 252-5210, Japan}
\altaffiltext{8}{%
Center for Planetary Science, Graduate School of Science, 
Kobe University, 7-1-48 Minatojima-Minamimachi,
Chuo-Ku, Kobe, Hyogo 650-0047, Japan}
\altaffiltext{9}{%
School of Medicine, Department of Basic Sciences,
University of Occupational and Environmental Health, 1-1 Iseigaoka,
Yahata, Kitakyusyu 807-8555, Japan}
\altaffiltext{10}{%
Planetary Exploration Research Center,
Chiba Institute of Technology, 2–17–1 Tsudanuma, Narashino,
Chiba, 275–0016, Japan}
\altaffiltext{11}{%
Research Center for the Early Universe, 
Graduate School of Science, The University of Tokyo, 
7-3-1 Hongo, Bunkyo-ku, Tokyo 113-0033, Japan}
\altaffiltext{12}{%
Astronomical Institute, Tohoku University, Sendai 980-8578, Japan}
\altaffiltext{13}{%
Division for the Establishment of Frontier Sciences, 
Organization for Advanced Studies, Tohoku University, Sendai 980-8577, Japan}
\altaffiltext{14}{%
Kavli Institute for the Physics and Mathematics of the Universe (WPI), 
The University of Tokyo Institutes for Advanced Study, 
The University of Tokyo, 5-1-5 Kashiwanoha, Kashiwa, Chiba 277-8583, Japan}
\altaffiltext{15}{%
National Astronomical Observatory of Japan, 
2-21-1 Osawa, Mitaka, Tokyo 181-8588, Japan}
\altaffiltext{16}{%
Department of Physics, Faculty of Science and Engineering,
Konan University, 8-9-1 Okamoto, Kobe, Hyogo 658-8501, Japan}
\altaffiltext{17}{%
The Hakubi Center/Astronomical Observatory, 
Graduate School of Science, Kyoto University, 
Kitashirakawa-oiwake-cho, Sakyo-ku, Kyoto 606-8502, Japan}


\maketitle

\begin{abstract}
We report the results of video observations of tiny
(diameter less than 100\,m) near-Earth objects (NEOs)
with Tomo-e Gozen on the Kiso 105\,cm Schmidt telescope.
A rotational period of a tiny asteroid reflects its dynamical history
and physical properties since smaller objects are sensitive to the YORP effect.
We carried out video observations of 60 tiny NEOs at 2\,fps from 2018 to 2021
and successfully derived the rotational periods and axial ratios of 32 NEOs
including 13 fast rotators with rotational periods less than 60\,s.
The fastest rotator found during our survey is 2020\,HS$_\mathsf{7}$
with a rotational period of 2.99\,s.
We statistically confirmed that 
there is a certain number of tiny fast rotators in the NEO population, 
which have been missed with any previous surveys.
We have discovered that the distribution of the tiny NEOs
in a diameter and rotational period (D-P) diagram is truncated
around a period of 10\,s.
The truncation with a flat-top shape is not explained well either by a
realistic tensile strength of NEOs or suppression of YORP by meteoroid impacts.
We propose that the dependence of the tangential YORP effect on the rotational
period potentially explains the observed pattern in the D-P diagram.
\end{abstract}


\section{Introduction}
As of March 2022, 28527 near-Earth objects (NEOs)
have been discovered by wide-field monitoring surveys such as
Catalina Sky Survey (CSS, \cite{Drake2009}),
Panoramic Survey Telescope 
and Rapid Response System (Pan-STARRS, \cite{Chambers2016}),
and Asteroid Terrestrial-impact Last Alert System\footnote{%
https://cneos.jpl.nasa.gov/stats/site\_all.html}
(ATLAS, \cite{Tonry2018}).
Most NEOs have their origins in the main belt
(e.g., \cite{Bottke2000b, Granvik2018}).
Asteroidal fragments are 
generated from collisional events in the main belt and their orbital elements 
are gradually changed by the Yarkovsky effect,
which is a thermal force caused by radiation from Sun
(e.g., \cite{Vokrouhlicky1998}, \,\yearcite{Vokrouhlicky2000}; \,\cite{Bottke2006}).
When the asteroids enter into orbital resonances with giant bodies, 
their orbits evolve to those of NEOs in a few Myr
(e.g., \cite{Gladman1997, Bottke2006}).
During the orbital evolution, the rotational states 
(i.e., rotational period and pole direction) 
of the object are changed by the
Yarkovsky-O'Keefe-Radzievskii-Paddack (YORP) effect,
which arises from asymmetricity 
of scattered sunlight and thermal radiation from its surface
(e.g., \cite{Rubincam2000};\,
Vokrouhlick{\'y} \& {\v{C}}apek\,\yearcite{Vokrouhlicky2002};\, 
{\v{C}}apek \& Vokrouhlick{\'{y}}\,\yearcite{Capek2004};\,
\cite{Bottke2006}).
The YORP effect caused by a recoil force normal to the surface, NYORP, is 
investigated intensively in previous studies.
Recently, the tangential YORP (TYORP) effect, which is caused by a recoil force 
parallel to the surface, was proposed by Golubov \& Kruguly\,(\yearcite{Golubov2012}).

Since the strength of the YORP effect increases with decreasing diameter,
smaller asteroids have been experienced larger change in the rotational states.
Thus, YORP is a dominant mechanism to change the rotational states of 
km-sized or smaller asteroids (Vokrouhlick{\'y} \& {\v{C}}apek\,\yearcite{Vokrouhlicky2002}).
The rotational acceleration by YORP leads to deformation or rotational fission 
of the asteroid due to a strong centrifugal force.
Because the YORP strength is also dependent on physical properties
such as shape and thermal conductivity,
the rotational period distribution of smaller objects probably reflects
the dynamical history and physical properties, which they have been undergone.

In general, it is difficult to constrain the rotational states of tiny asteroids 
due to 
limited observational windows (hours to days),
fast rotation (less than a minute),
and large apparent motion on the sky (a few $\mathrm{arcsec\,\,s^{-1}}$).
Observations with exposure times sufficiently shorter than
their rotational periods are required.
The shorter exposure times are effective to suppress the trailing sensitivity loss effect, 
which is the degradation of a surface brightness of a moving object on an image \citep{Zhai2014}. 

\begin{figure}
  \includegraphics[width=80mm]{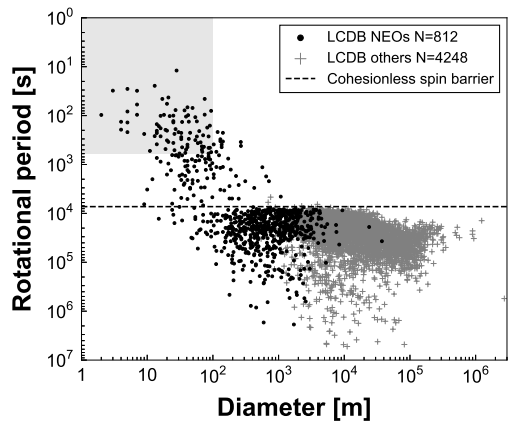}
  \caption{%
    Diameter and rotational period relation of the objects in the LCDB 
    \citep{Warner2009} as of June 2021. 
    NEOs and other objects (main belt and trans-Neptunian objects) are 
    presented in filled circles and plus signs, respectively.
    A cohesionless spin barrier assuming a typical density of S-type asteroids of
    2.67 g\,cm$^{\text{-3}}$ \citep{Yeomans2000} is shown by a dashed line.
    Tiny ($D \leq 100\,\mathrm{m}$) and fast ($P \leq 10\,\mathrm{min}$)
    biased region is shown as a gray shaded area.
    }\label{fig:LCDB}
\end{figure}

The Asteroid Light Curve Database (LCDB, \cite{Warner2009})
has thousands of rotational periods of minor planets.
The diameter and rotational period relation 
(hereinafter referred to as D-P relation) is shown in figure \ref{fig:LCDB}.
As of June 2021, rotational periods of 5,060 objects are estimated 
with high accuracy (the quality code $U$ in \cite{Warner2009} is 
\texttt{3} or \texttt{3-}).
For asteroids larger than 200\,m in diameter, 
the rotation period distribution is truncated around 2 hours. 
This clear structure is called the cohesionless 
spin barrier and indicates most of the larger asteroids are rubble-piles
(Pravec \& Harris\,\yearcite{Pravec2000}).
It is possible to constrain physical properties of asteroids 
smaller than 200\,m in diameter from the D-P relation as same as the larger asteroids.
However, there is a smaller number of asteroids for which 
the rotational period has been reported so far.

The LCDB contains the observational results of 
Mission Accessible Near-Earth Objects Survey 
(MANOS), which obtained more than 300 light curves of small NEOs 
with the mean absolute magnitude of about 24\,mag 
using large and medium aperture telescopes (\cite{Thirouin2016},\,\yearcite{Thirouin2018}).
Although MANOS successfully derived the rotational periods of NEOs with high accuracy, 
the main motivation of the survey is not to detect fast rotators, 
but to characterize mission accessible NEOs.
Due to a relatively long exposure time (1--300\,s), 
the survey possibly undetected the very fast rotations. 
Systematic high-speed observations are required
to correctly derive shorter rotational periods and obtain an unbiased 
D-P relation of tiny NEOs.

In this paper, 
we report the results of imaging observations at 2\,fps
of 60 tiny NEOs with the wide-field CMOS camera Tomo-e Gozen.
The observed NEOs are smaller than 100\,m in diameter 
and their mean diameter is 20\,m.
The aims of this study are
to obtain an unbiased D-P relation by video observations 
and to reveal dynamical histories 
and physical properties of tiny NEOs.
Observations and data reduction are described in section 2.
The results are compared with previous studies in section 3.
In section 4, the D-P relation of the tiny NEOs obtained in this study
is discussed taking into account the spin acceleration by YORP.

\section{Observations and data reduction}

\subsection{Observations}
We conducted photometric observations at 2\,fps
with the wide-field CMOS camera Tomo-e Gozen \citep{Sako2018}.
Tomo-e Gozen is a wide-field high-speed camera mounted on the 105\,cm 
Schmidt telescope at Kiso Observatory (Minor Planet Center code 381) in Nagano, Japan.
The field of view is 20.7\,square degrees covered by 84 chips of CMOS sensors
without photometric filters.
A timestamp of each image data is GPS-synchronized and has a time accuracy of 0.2\,milliseconds.
We have performed 2\,fps all-sky survey observations with Tomo-e Gozen since 2019.
Data accumulated each night amount to 30\,TB, 
from which various types of transients
such as supernovae and tiny NEOs are searched for.  
Tomo-e Gozen has discovered 32 NEOs
from the survey data in real time
from March 2019 to October 2021
with the fast-moving object pipeline using machine-learning technique 
(Ohsawa et al. in prep.).
The algorithms used in the pipeline are partly described in \citet{Ohsawa2021}.

We have obtained light curves of 60 NEOs from May 2018 to October 2021.
Nominal criteria for target selection are that a
$V$-band apparent magnitude ($V$) is smaller than 17 
and an absolute magnitude ($H$) is larger than 22.5.
We referred $V$ and $H$ from the website of International Astronomical Union 
Minor Planet Center\footnote{https://minorplanetcenter.net/}
to make observation plans.
We call our selected samples hereinafter the Tomo-e NEOs, which are listed in table \ref{tab:obsneo1}.
The observation specifications are referenced from 
NASA JPL/HORIZONS\footnote{https://ssd.jpl.nasa.gov/?horizons} with 
\texttt{astroquery.jplhorizons} \citep{Ginsburg2019}.
The Tomo-e NEOs consists of 37 NEOs discovered by other facilities and 23 NEOs discovered by Tomo-e Gozen itself.
The $V$-band magnitude of 17 corresponds to a 5-sigma limiting magnitude 
in 2\,fps video observations with Tomo-e Gozen.
An asteroid diameter ($D$) in table \ref{tab:obsneo1} is derived from $H$ using the equation (Fowler\,\&\,Chillemi \yearcite{Fowler1992};\, Pravec \& Harris\,\yearcite{Pravec2007}):
\begin{equation}
    D=\frac{1.329\times10^6}{\sqrt{p_V}}\times10^{-\frac{H}{5}}\,\,\mathrm{m},
    \label{eq:D}
\end{equation}
where $p_V$ is a geometric albedo in $V$-band.

\begin{figure}
  \includegraphics[width=8cm]{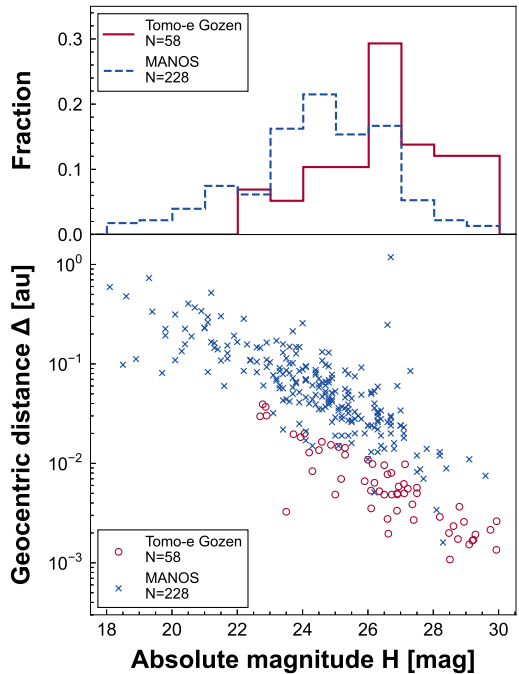} 
    \caption{%
    (upper panel) Fractional distribution of absolute magnitudes.
    Tomo-e NEOs and MANOS NEOs are illustrated by solid and dashed lines, respectively.
    (lower panel) 
    Absolute magnitude versus geocentric distance 
    of the Tomo-e NEOs (circles)
    and the MANOS NEOs (crosses) at the observation times.
    The absolute magnitude and the geocentric distances are referenced from
    NASA JPL/HORIZONS 
    as of 2021-12-27 (UTC).
    NEO candidates, TMG0042 and TMG0049, are not included in this figure.
    }\label{fig:MANOStomoe}
\end{figure}

In this paper, we assume that $p_V$ is 0.20,
which is a typical value for S-type asteroids, as used in LCDB. 
The absolute magnitude of 22.5 corresponds to 94\,m in diameter.
Since a median rotational period of NEOs in LCDB satisfying
the quality code $U$ of \texttt{3} or \texttt{3-} 
and $H$ smaller than 22.5 is about 9 min,
we set a nominal duration of observation as 20\,min.
A mean absolute magnitude of our samples is 26.6 corresponding to 14\,m in diameter.
As shown in figure \ref{fig:MANOStomoe}, 
the peak of the distribution ($H \sim 26$) is smaller than the peak of the targets observed 
by MANOS($H \sim 24$), hereinafter referred to as the MANOS NEOs.

The Tomo-e NEOs were typically located at a few lunar distances from Earth when observed.
A typical angular velocity was about a few $\mathrm{arcsec\,\,s^{-1}}$.
Most of the Tomo-e NEOs were discovered a few hours or a few days before our observations,
except for $2010\,\mathrm{WC_9}$, $2011\,\mathrm{DW}$, and $2017\,\mathrm{WJ_{16}}$.
TMG0042 and TMG0049 are NEO candidates discovered by Tomo-e Gozen.
Due to the limited number of follow-up observations,
provisional designations for the two objects were not served from Minor Planet Center.

To obtain the light curve of the NEO,
we used a single sensor of Tomo-e Gozen with a field of view of 
$\ang{;39.7;}\times\ang{;22.4;}$ 
and a pixel scale of 1.189 arcsec.
Sidereal tracking and re-pointing were performed to follow the fast-moving NEOs.
All of the Tomo-e NEOs except for $2018\,\mathrm{LV_3}$ were observed at 2\,fps.
The light curve of $2018\,\mathrm{LV_3}$ was obtained at 0.2\,fps 
as an experimental observation.

   \begin{longtable}{
     p{16mm}p{9mm}p{7mm}p{5mm}p{30mm}
     p{5mm}p{7mm}p{15mm}p{7mm}p{7mm}
     p{7mm}p{25mm}}
     \caption{Summary of observations.$^{*}$.}
     \hline
         Object                & Dyn.       & \multicolumn{1}{c}{$H$}     & \multicolumn{1}{c}{$D^{\dag}$}   & Obs. Date  & $T$   &   $V$  & Vel.        & $\alpha$   & $\Delta$  & $r$  & Note  \\ 
                               & Class      & \multicolumn{1}{c}{(mag)}    & \multicolumn{1}{c}{(m)}                         & (UTC)      & (min)   &  (mag)  &  ($\mathrm{arcsec\,\,s^{-1}}$) & ($^\circ$) & (au)    & (au) & \\
     \hline
     \endfirsthead
     \hline
    Object                & Dyn.       &  \multicolumn{1}{c}{$H$}     & \multicolumn{1}{c}{$D^{\dag}$}   & Obs. Date  & $T$   &   $V$  & Vel.        & $\alpha$   & $\Delta$  & $r$  & Note  \\ 
                               & Class      & \multicolumn{1}{c}{(mag)}    & \multicolumn{1}{c}{(m)}                        & (UTC)      & (min)   &  (mag)  &  ($\mathrm{arcsec\,\,s^{-1}}$) & ($^\circ$) & (au)    & (au) & \\
     \hline
     \endhead
     \hline
     \endfoot
     \label{tab:obsneo1}
    \multicolumn{12}{l}{\footnotesize{
    $^{*}$ Dynamical class (Dyn. Class) and absolute magnitude ($H$) are from 
      NASA JPL/HORIZONS
     as of 2022-1-9 (UTC).
    Observation}}
    \\ \multicolumn{12}{l}{\footnotesize{
    starting time in UTC (Obs. Date) and duration time of observation ($T$) 
    for each object are listed.
    $V$-band apparent magnitude ($V$), 
    }}
    \\ \multicolumn{12}{l}{\footnotesize{
    angular rate of change in apparent RA and DEC (Vel.),
    phase angle ($\alpha$),
    distance between NEO and observer ($\Delta$),
    and distance between 
    }}
    \\ \multicolumn{12}{l}{\footnotesize{
    Sun and NEO ($r$) at the observation time 
    are also from NASA JPL/HORIZONS
     as of 2022-1-9 (UTC).
    }}\\ \multicolumn{12}{l}{\footnotesize{
    $^{\dag}$ Diameter ($D$) is derived from $H$ 
    assuming geometric albedo in $V$-band of 0.20.
    }}
    \\ \multicolumn{12}{l}{\footnotesize{
    $^{\ddag}$ Dyn. Class and $H$ of the NEO candidates are derived from orbits
    determined with the Tomo-e Gozen data using Find\_Orb
    }}
    \\ \multicolumn{12}{l}{\footnotesize{
    (\url{https://www.projectpluto.com/fo.htm}).
    }}
    \endlastfoot$2010\,\mathrm{WC_{9}}$ & Apollo &\multicolumn{1}{c}{23.5}&\multicolumn{1}{c}{59}&2018-05-15 12:19:01 & 14.0 & 12.3 &2.3 & 28.0 &0.0033 & 1.0138 &  \\
$2011\,\mathrm{DW}$ & Aten &\multicolumn{1}{c}{22.9}&\multicolumn{1}{c}{79}&2021-02-28 15:15:36 & 20.0 & 16.4 &0.5 & 10.0 &0.0370 & 1.0271 &  \\
$2017\,\mathrm{WJ_{16}}$ & Aten &\multicolumn{1}{c}{24.5}&\multicolumn{1}{c}{37}&2020-11-23 17:42:48 & 16.0 & 16.7 &0.4 & 37.1 &0.0136 & 0.9982 &  \\
  &  &  &  &2020-11-25 13:30:59 & 60.0 & 16.7 &0.3 & 34.9 &0.0141 & 0.9986 &  \\
$2018\,\mathrm{LV_{3}}$ & Apollo &\multicolumn{1}{c}{26.5}&\multicolumn{1}{c}{15}&2018-06-13 15:18:08 & 100.0 & 17.9 &0.4 & 33.6 &0.0096 & 1.0236 & 5 s exposure\\
$2018\,\mathrm{UD_{3}}$ & Apollo &\multicolumn{1}{c}{26.2}&\multicolumn{1}{c}{17}&2018-11-01 13:47:46 & 36.5 & 17.1 &1.0 & 50.3 &0.0064 & 0.9966 &  \\
$2019\,\mathrm{BE_{5}}$ & Aten &\multicolumn{1}{c}{25.1}&\multicolumn{1}{c}{28}&2019-02-01 09:58:47 & 114.0 & 17.5 &0.7 & 39.1 &0.0146 & 0.9967 &  \\
$2020\,\mathrm{EO}$ & Apollo &\multicolumn{1}{c}{25.9}&\multicolumn{1}{c}{20}&2020-03-13 10:47:05 & 13.0 & 16.3 &1.1 & 29.6 &0.0066 & 0.9998 &  \\
$2020\,\mathrm{FA_{2}}$ & Apollo &\multicolumn{1}{c}{27.5}&\multicolumn{1}{c}{9}&2020-03-18 16:53:30 & 44.0 & 17.4 &1.3 & 22.5 &0.0057 & 1.0008 &  \\
$2020\,\mathrm{FL_{2}}$ & Apollo &\multicolumn{1}{c}{26.1}&\multicolumn{1}{c}{18}&2020-03-22 13:28:48 & 14.0 & 14.5 &2.1 & 11.7 &0.0035 & 1.0001 &  \\
$2020\,\mathrm{GY_{1}}$ & Apollo &\multicolumn{1}{c}{26.6}&\multicolumn{1}{c}{14}&2020-04-05 17:16:28 & 14.0 & 14.9 &1.3 & 24.0 &0.0028 & 1.0032 &  \\
$2020\,\mathrm{HK_{3}}$ & Apollo &\multicolumn{1}{c}{24.2}&\multicolumn{1}{c}{43}&2020-04-30 13:41:54 & 13.0 & 17.2 &1.6 & 69.2 &0.0129 & 1.0120 &  \\
$2020\,\mathrm{HS_{7}}$ & Apollo &\multicolumn{1}{c}{29.1}&\multicolumn{1}{c}{4}&2020-04-28 14:36:31 & 14.0 & 16.0 &3.0 & 19.1 &0.0015 & 1.0085 &  \\
  &  &  &  &2020-04-28 16:24:41 & 9.0 & 15.1 &8.4 & 25.7 &0.0009 & 1.0079 &  \\
$2020\,\mathrm{HT_{7}}$ & Apollo &\multicolumn{1}{c}{26.9}&\multicolumn{1}{c}{12}&2020-04-27 16:40:37 & 13.0 & 16.9 &1.1 & 39.5 &0.0049 & 1.0105 &  \\
$2020\,\mathrm{HU_{3}}$ & Apollo &\multicolumn{1}{c}{26.0}&\multicolumn{1}{c}{19}&2020-04-21 16:23:06 & 6.0 & 17.6 &1.3 & 32.0 &0.0109 & 1.0144 &  \\
$2020\,\mathrm{PW_{2}}$ & Apollo &\multicolumn{1}{c}{28.8}&\multicolumn{1}{c}{5}&2020-08-14 16:46:48 & 24.0 & 17.8 &2.0 & 24.8 &0.0037 & 1.0162 & crowded field\\
$2020\,\mathrm{PY_{2}}$ & Apollo &\multicolumn{1}{c}{26.5}&\multicolumn{1}{c}{15}&2020-08-20 12:50:56 & 23.0 & 15.7 &2.3 & 12.1 &0.0049 & 1.0165 &  \\
$2020\,\mathrm{QW}$ & Apollo &\multicolumn{1}{c}{25.3}&\multicolumn{1}{c}{26}&2020-08-17 16:10:51 & 20.0 & 18.3 &0.9 & 73.4 &0.0122 & 1.0158 & crowded field\\
$2020\,\mathrm{TD_{8}}$ & Apollo &\multicolumn{1}{c}{26.9}&\multicolumn{1}{c}{12}&2020-10-26 16:59:35 & 17.0 & 17.2 &1.0 & 48.3 &0.0050 & 0.9972 &  \\
$2020\,\mathrm{TE_{6}}$ & Apollo &\multicolumn{1}{c}{27.4}&\multicolumn{1}{c}{10}&2020-10-18 10:30:50 & 19.0 & 16.8 &3.2 & 62.9 &0.0027 & 0.9975 &  \\
$2020\,\mathrm{TS_{1}}$ & Aten &\multicolumn{1}{c}{29.2}&\multicolumn{1}{c}{4}&2020-10-12 10:13:04 & 9.0 & 16.8 &3.6 & 37.4 &0.0017 & 0.9993 &  \\
$2020\,\mathrm{UQ_{6}}$ & Apollo &\multicolumn{1}{c}{22.7}&\multicolumn{1}{c}{86}&2020-10-28 17:30:35 & 17.0 & 16.0 &0.4 & 16.0 &0.0297 & 1.0219 &  \\
$2020\,\mathrm{VF_{4}}$ & Apollo &\multicolumn{1}{c}{26.6}&\multicolumn{1}{c}{14}&2020-11-13 16:56:23 & 20.0 & 17.0 &2.0 & 18.5 &0.0078 & 0.9968 &  \\
$2020\,\mathrm{VH_{5}}$ & Apollo &\multicolumn{1}{c}{29.2}&\multicolumn{1}{c}{4}&2020-11-13 17:33:04 & 21.0 & 15.9 &5.4 & 7.7 &0.0017 & 0.9912 &  \\
$2020\,\mathrm{VJ_{1}}$ & Apollo &\multicolumn{1}{c}{26.7}&\multicolumn{1}{c}{13}&2020-11-09 15:23:30 & 20.0 & 16.6 &3.1 & 34.1 &0.0049 & 0.9944 &  \\
$2020\,\mathrm{VR_{1}}$ & Apollo &\multicolumn{1}{c}{28.9}&\multicolumn{1}{c}{5}&2020-11-09 15:44:37 & 16.0 & 17.5 &6.4 & 36.1 &0.0026 & 0.9925 &  \\
$2020\,\mathrm{VZ_{6}}$ & Apollo &\multicolumn{1}{c}{25.0}&\multicolumn{1}{c}{30}&2020-12-02 14:07:02 & 14.0 & 14.7 &1.1 & 31.0 &0.0049 & 0.9900 &  \\
$2020\,\mathrm{XH}$ & Apollo &\multicolumn{1}{c}{24.6}&\multicolumn{1}{c}{36}&2020-12-05 16:35:39 & 17.0 & 16.8 &0.6 & 24.2 &0.0165 & 1.0004 & crowded field\\
$2020\,\mathrm{XH_{1}}$ & Apollo &\multicolumn{1}{c}{22.9}&\multicolumn{1}{c}{78}&2020-12-08 12:56:09 & 20.0 & 16.7 &0.4 & 30.9 &0.0302 & 1.0108 &  \\
$2020\,\mathrm{XQ_{2}}$ & Apollo &\multicolumn{1}{c}{22.8}&\multicolumn{1}{c}{83}&2020-12-09 15:00:39 & 20.0 & 16.6 &0.8 & 15.0 &0.0393 & 1.0228 &  \\
$2020\,\mathrm{XX_{3}}$ & Apollo &\multicolumn{1}{c}{28.5}&\multicolumn{1}{c}{6}&2020-12-17 14:09:52 & 18.0 & 16.5 &1.0 & 38.0 &0.0020 & 0.9856 &  \\
$2020\,\mathrm{XY_{4}}$ & Aten &\multicolumn{1}{c}{26.9}&\multicolumn{1}{c}{12}&2020-12-20 11:10:51 & 20.0 & 17.3 &2.1 & 40.3 &0.0059 & 0.9883 & thin cloud\\
$2020\,\mathrm{YJ_{2}}$ & Apollo &\multicolumn{1}{c}{27.4}&\multicolumn{1}{c}{10}&2020-12-21 14:09:09 & 20.0 & 16.7 &2.3 & 35.7 &0.0039 & 0.9869 & crowded field\\
$2021\,\mathrm{AT_{5}}$ & Apollo &\multicolumn{1}{c}{27.5}&\multicolumn{1}{c}{9}&2021-01-13 13:40:09 & 10.0 & 16.9 &1.8 & 17.9 &0.0050 & 0.9884 & crowded field\\
$2021\,\mathrm{BC}$ & Aten &\multicolumn{1}{c}{24.3}&\multicolumn{1}{c}{41}&2021-01-21 10:55:35 & 18.0 & 15.9 &1.8 & 55.7 &0.0083 & 0.9888 &  \\
$2021\,\mathrm{CA_{6}}$ & Apollo &\multicolumn{1}{c}{28.5}&\multicolumn{1}{c}{6}&2021-02-13 16:03:04 & 22.0 & 16.0 &8.1 & 66.3 &0.0011 & 0.9879 &  \\
$2021\,\mathrm{CC_{7}}$ & Apollo &\multicolumn{1}{c}{29.8}&\multicolumn{1}{c}{3}&2021-02-12 18:06:58 & 11.0 & 17.1 &2.6 & 11.5 &0.0021 & 0.9894 &  \\
$2021\,\mathrm{CG}$ & Apollo &\multicolumn{1}{c}{26.1}&\multicolumn{1}{c}{18}&2021-02-06 15:18:41 & 20.0 & 17.0 &1.1 & 16.7 &0.0098 & 0.9957 &  \\
$2021\,\mathrm{CO}$ & Apollo &\multicolumn{1}{c}{25.3}&\multicolumn{1}{c}{26}&2021-02-09 12:12:05 & 21.0 & 16.6 &0.2 & 7.3 &0.0143 & 1.0009 &  \\
$2021\,\mathrm{DW_{1}}$ & Apollo &\multicolumn{1}{c}{25.2}&\multicolumn{1}{c}{27}&2021-03-02 11:14:22 & 2.0 & 16.2 &0.5 & 49.2 &0.0070 & 0.9957 & crowded field\\
$2021\,\mathrm{EM_{4}}$ & Apollo &\multicolumn{1}{c}{27.1}&\multicolumn{1}{c}{11}&2021-03-18 16:04:37 & 20.0 & 16.8 &1.5 & 26.6 &0.0050 & 0.9999 &  \\
$2021\,\mathrm{EQ_{3}}$ & Apollo &\multicolumn{1}{c}{26.1}&\multicolumn{1}{c}{18}&2021-03-15 11:07:03 & 14.0 & 16.6 &1.0 & 49.6 &0.0053 & 0.9980 &  \\
$2021\,\mathrm{ET_{4}}$ & Apollo &\multicolumn{1}{c}{23.9}&\multicolumn{1}{c}{48}&2021-03-16 14:20:17 & 18.0 & 17.0 &0.7 & 45.1 &0.0184 & 1.0078 &  \\
$2021\,\mathrm{EX_{1}}$ & Apollo &\multicolumn{1}{c}{24.9}&\multicolumn{1}{c}{32}&2021-03-08 12:53:11 & 20.0 & 16.9 &0.5 & 22.3 &0.0153 & 1.0069 &  \\
$2021\,\mathrm{FH}$ & Apollo &\multicolumn{1}{c}{26.7}&\multicolumn{1}{c}{13}&2021-03-22 13:13:51 & 20.0 & 17.3 &0.3 & 21.5 &0.0080 & 1.0040 & thin cloud\\
$2021\,\mathrm{GD_{5}}$ & Aten &\multicolumn{1}{c}{27.1}&\multicolumn{1}{c}{11}&2021-04-08 15:38:36 & 20.0 & 18.2 &2.7 & 24.3 &0.0098 & 1.0104 &  \\
$2021\,\mathrm{GQ_{10}}$ & Apollo &\multicolumn{1}{c}{26.6}&\multicolumn{1}{c}{14}&2021-04-14 16:03:02 & 20.0 & 15.4 &2.8 & 65.7 &0.0020 & 1.0040 &  \\
$2021\,\mathrm{GT_{3}}$ & Apollo &\multicolumn{1}{c}{26.4}&\multicolumn{1}{c}{16}&2021-04-10 13:03:32 & 20.0 & 15.7 &2.1 & 11.5 &0.0052 & 1.0071 &  \\
$2021\,\mathrm{JB_{6}}$ & Apollo &\multicolumn{1}{c}{28.8}&\multicolumn{1}{c}{5}&2021-05-13 15:11:10 & 20.0 & 16.8 &2.7 & 47.3 &0.0017 & 1.0118 &  \\
$2021\,\mathrm{KN_{2}}$ & Apollo &\multicolumn{1}{c}{28.6}&\multicolumn{1}{c}{6}&2021-05-30 16:53:29 & 14.0 & 17.1 &2.3 & 41.2 &0.0023 & 1.0156 &  \\
$2021\,\mathrm{KQ_{2}}$ & Aten &\multicolumn{1}{c}{29.9}&\multicolumn{1}{c}{3}&2021-05-31 16:27:10 & 20.0 & 17.2 &2.9 & 39.1 &0.0014 & 1.0150 &  \\
$2021\,\mathrm{RB_{1}}$ & Amor &\multicolumn{1}{c}{24.1}&\multicolumn{1}{c}{46}&2021-09-06 13:16:10 & 20.0 & 16.8 &1.1 & 26.5 &0.0199 & 1.0258 &  \\
$2021\,\mathrm{RX_{5}}$ & Apollo &\multicolumn{1}{c}{23.7}&\multicolumn{1}{c}{54}&2021-09-15 15:25:56 & 7.0 & 16.6 &0.5 & 33.2 &0.0196 & 1.0219 &  \\
$2021\,\mathrm{TG_{1}}$ & Apollo &\multicolumn{1}{c}{28.2}&\multicolumn{1}{c}{7}&2021-10-03 13:22:32 & 20.0 & 17.1 &2.5 & 39.4 &0.0029 & 1.0028 &  \\
$2021\,\mathrm{TL_{14}}$ & Apollo &\multicolumn{1}{c}{26.9}&\multicolumn{1}{c}{12}&2021-10-14 14:45:00 & 21.0 & 15.7 &2.4 & 27.8 &0.0033 & 1.0004 &  \\
$2021\,\mathrm{TQ_{3}}$ & Atira &\multicolumn{1}{c}{27.1}&\multicolumn{1}{c}{11}&2021-10-06 16:25:11 & 20.0 & 17.1 &1.4 & 21.3 &0.0062 & 1.0055 &  \\
$2021\,\mathrm{TQ_{4}}$ & Apollo &\multicolumn{1}{c}{29.9}&\multicolumn{1}{c}{3}&2021-10-06 16:54:27 & 3.0 & 17.3 &3.1 & 2.4 &0.0026 & 1.0023 &  \\
$2021\,\mathrm{TY_{14}}$ & Apollo &\multicolumn{1}{c}{27.2}&\multicolumn{1}{c}{11}&2021-10-15 11:56:39 & 20.0 & 17.0 &1.8 & 22.2 &0.0056 & 1.0023 &  \\
$2021\,\mathrm{UF_{12}}$ & Apollo &\multicolumn{1}{c}{29.3}&\multicolumn{1}{c}{4}&2021-10-29 14:46:18 & 8.0 & 16.5 &9.2 & 14.0 &0.0019 & 0.9951 &  \\
TMG0042 & Apollo$^{\ddag}$&\multicolumn{1}{c}{28.5$^{\ddag}$}&\multicolumn{1}{c}{6}&2021-04-10 16:16:17 & 20.0 & - & - & - & - & - & NEO candidate\\
TMG0049 & Apollo$^{\ddag}$&\multicolumn{1}{c}{30.0$^{\ddag}$}&\multicolumn{1}{c}{3}&2021-05-30 15:30:09 & 16.0 & - & - & - & - & - & NEO candidate\\

    \hline
    \end{longtable}

\subsection{Data reduction}
\subsubsection{Photometry}
Observations are composed of a series of video data, that were typically 1 minute in length.
The video data were compiled into cube FITS files.
After bias and dark subtraction and flat-field correction,
standard circle aperture photometry was performed on a target and reference stars
in each frame using the SExtractor-based python package \texttt{sep}
(Bertin \& Arnouts\,\yearcite{Bertin1996};\,\cite{Barbary2015}).
Since the elongations of the NEOs were negligible, 
we applied the standard aperture photometry method.
The aperture radius was set to 2 to 3 times larger than the full width at half maximum (FWHM)
of the point spread function (PSF) of reference stars, which was typically 3 to 5 arcsec.
We determined the FWHM of the stellar PSF in the first frame of the cube,
and then conducted the photometry of the objects in each frame.
Sometimes the target was too faint to be detected possibly by the brightness variation of the target.
In such cases, we set the aperture at the expected positions
interpolated from the positions in adjacent frames and performed forced photometry.

We used the $G$-band magnitude of Gaia DR2 catalog 
as brightness references since the spectral response of Tomo-e Gozen 
(350 to 950\,nm, \cite{Kojima2018})
is similar to that of the $G$-band of Gaia (330 to 1050 nm, \cite{Gaia2018}).
The difference in the spectral responses may affect the mean apparent magnitudes
of the NEOs, 
but the rotational period is not affected by the spectral response
and the effect on the amplitudes are negligible.
The discussion in this paper is not affected.

The $G$-band magnitude of a NEO, $m_{G}$, on each frame was derived as follows:
\begin{equation}
    m_G = -2.5 \log_{10}F + Z
    \label{eq:mag},
\end{equation}
where $F$ is a total flux in the aperture, $Z$ is the magnitude zero point of the frame. 
Stars with $G$-band magnitudes $10 < m_G < 15$ and 
broad-band colors $-1 < G_{\mathrm{BP}} - G_{\mathrm{BP}} < 1$,
typically 20--30, were used to calculate 
the magnitude zero points and the median value of the zero points was used as $Z$.
An uncertainty of $Z$ was estimated from the median absolute deviation of the zero points.
The photometric error of the NEO consists of 
the background noise, the Poisson noise, and the uncertainty of $Z$.

The observed $G$-band magnitudes were 
converted to reduced magnitudes
with distance between Sun and NEO ($r$) 
and NEO and observer ($\Delta$) at the time of observations.
The phase angle correction and 
the light-travel time correction were done 
to obtain the corrected light curves.

\subsubsection{Periodic analysis}\label{periana}
We used the Lomb-Scargle technique
to estimate rotational periods from non-evenly sampled data
\citep{Lomb1976, Scargle1982, VanderPlas2018}.
Fitting models are given by the following form:
\begin{equation}
    y_{\mathrm{model}}(f, t) = c_0+\sum_{i=1}^{n}\{ S_i\sin(2\pi ift) + C_i\cos (2\pi ift)\},
    \label{eq:LS}
\end{equation}
where $f$ is a frequency, 
$c_0$ denotes the average brightness,
$n$ is the number of harmonics,
and $S_i$ and $C_i$ are the Fourier coefficients of the $i$-th harmonics.
The normalized residual $\chi^2$ was calculated as
\begin{equation}
    \chi^2(f) = \sum_{j=1}^{n_{\mathrm{obs}}}\left(\frac{y_{\mathrm{model}}(f, t_j) - y_{\mathrm{obs},j}}{y_{\mathrm{err},j}}\right)^2
    \label{eq:chi2},
\end{equation}
where 
$n_{\mathrm{obs}}$ is the number of observation data,
$t_j$ is an observation time of the $j$-th sample, 
$y_{\mathrm{obs},j}$ and $y_{\mathrm{err},j}$ are the $j$-th measured brightness 
and its uncertainty, respectively.
We calculated the Lomb-Scargle periodogram $P_{\mathrm{LS}}$ as
\begin{equation}
    P_{\mathrm{LS}}(f)=\frac{\chi_0^2 - \chi^2(f)}{2},
\label{eq:lombP}
\end{equation}
where $\chi_0^2$ is a $\chi^2(f)$ for a constant fitting model 
where $S_i$ and $C_i$ are set to zero for all $i$.

The significance of a peak in a periodogram is evaluated by 
calculating a confidence level against the null hypothesis.
Assuming that the data consists of pure Gaussian noise,
$2 P_\mathrm{LS}$ follows a $\chi^2$ distribution with two degrees of freedom when $n=1$.
Thus,
\begin{equation}
    p_{\mathrm{single}}(z) = 1 - e^{-z}
\end{equation}
expresses a cumulative probability that $P_\mathrm{LS}$ is less than $z$ at each frequency.
We assumed that the frequencies are independent each other 
and defined an effective number of frequencies as
\begin{equation}
    N_{\mathrm{eff}}=f_{\mathrm{max}}T,
\end{equation}
where $f_{\mathrm{max}}$ is a maximum frequency to be considered. 
A false alarm probability, $FAP(z)$, is calculated as follows:
\begin{equation}
    FAP(z)=1-p_{\mathrm{single}}(z)^{N_{\mathrm{eff}}}. \label{eq:fal}
\end{equation}
We calculated a 99.9\% confidence level in each periodogram.
We derived a candidate of a rotational period from the highest peak 
of $P_\mathrm{LS}$ larger than 99.9\% confidence level.

For optimal determination of the number of harmonics $n$,
we used the Akaike Information Criterion ($AIC$, \cite{Akaike1974}).
$AIC$ indicates the trade-off between the goodness of fit and the simplicity of the model. 
$AIC$ is calculated as
\begin{eqnarray}
AIC &= -\ln L + 2(2n+1) \nonumber \\
&= \frac{1}{2}\sum_{j=1}^{n_{\mathrm{obs}}} \left(\frac{y_{\mathrm{model}}(f,t_j)-y_{\mathrm{obs},j}}{y_{\mathrm{err},j}}\right)^2 \nonumber \\
& +\ln{\prod_{j=1}^{n_{\mathrm{obs}}}{y_{\mathrm{err},j}}} + \frac{n_{\mathrm{obs}}}{2}\ln{2\pi} + 2(2n+1),
\end{eqnarray}
where $L$ is the likelihood of the parameters.
We adopted $n$ of each NEO for which $AIC$ value is the minimum.

\begin{figure}
    \includegraphics[width=80mm]{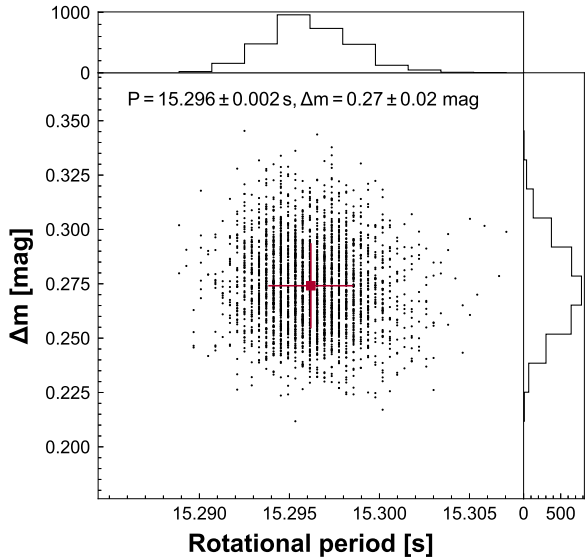}
    \caption{%
         Scatter plot of the rotational periods and the light curve amplitudes 
         of 3000 model curves of $2021\,\mathrm{CG}$.
         Histograms present the marginal distributions of the periods 
         and the amplitudes, respectively.
         The derived period and amplitude are presented by a square symbol
         with a cross that indicates the standard deviations of the period and the amplitude.
      }\label{fig:MC2021CG}
\end{figure}

The uncertainty of the rotational period and the light curve amplitude were estimated using 
the Monte Carlo method.
We created 3000 light curves for each NEO 
by randomly resampling the data assuming each observed data follows a normal distribution 
whose standard deviation is a photometric error.
We performed the same analyses above for 3000 light curves and obtained 3000 sets of Fourier coefficients in equation (\ref{eq:LS}).
We calculated the 3000 periods and the 3000 amplitude with the corresponding peak frequencies for each light curve.
As an example, an analysis result of $2021\,\mathrm{CG}$ is shown in figure \ref{fig:MC2021CG}.
We adopted the standard deviations as the uncertainties of the period and the amplitude.
If the estimated standard deviation was larger than 5\,\% of the rotational period,
we judged the derived rotational period was suspicious 
and the correct rotational period was not derived.
We used the Fourier coefficients of which the rotational period and the light curve amplitude 
are the closest to the average values to plot a typical model curve.

There are several candidates of non-principal axis rotators (i.e., tumbler) in the Tomo-e NEOs.
The tumbler is in a excited state and its light curve is 
complicated (e.g., \cite{Paolicchi2002, Pravec2005}).
Therefore, the periodograms of the tumblers show additional peaks 
which are not aliases of the highest peak.
We defined such objects with multiple peaks as tumbler candidates in this paper.

\section{Results}\label{sec3}
We successfully derived the rotational periods of 32 NEOs.
Results of the analysis are summarized in table \ref{tab:resneo1}.
The rotational periods of 11 out of the 32 are reported in previous studies:
$2019\,\mathrm{BE_5}$
\citep{Warner2009},
$2020\,\mathrm{TD_8}$, $2020\,\mathrm{UQ_6}$, $2020\,\mathrm{VZ_6}$, $2020\,\mathrm{XX_3}$
\citep{Birtwhistle2021a}, 
$2021\,\mathrm{EX_1}$, $2021\,\mathrm{FH}$
\citep{Birtwhistle2021b}, 
$2021\,\mathrm{KN_2}$, $2021\,\mathrm{JB_6}$, $2021\,\mathrm{GQ_{10}}$ 
\citep{Birtwhistle2021c},
and $2021\,\mathrm{DW_1}$ 
\citep{Kwiatkowski2021}.
All our results are consistent with the reported values.
The periodogram of $2021\,\mathrm{FH}$ has a prominent peak but its significance level is lower than 99.9\%.
We considered that the peak of $2021\,\mathrm{FH}$ is reliable 
since the peak frequency corresponds to the rotational period (63.4\,s) reported by \citet{Birtwhistle2021b}.

The rotational periods of 18 objects were not derived due to small amplitudes.
These objects may have 
axisymmetric shapes, 
rotational periods longer than the duration of observation, 
rotational periods shorter than the exposure time, 
or rotational axes parallel to the line of sight.
When a light curve shows a clear brightness variation but whole cycles of rotation were not obtained,
we adopted the duration times of the observations as lower limits of rotational periods.

We found 10 candidates of tumbler:
$2010\,\mathrm{WC_9}$, 
$2017\,\mathrm{WJ_{16}}$, 
$2020\,\mathrm{TE_{6}}$, 
$2021\,\mathrm{CO}$, 
$2020\,\mathrm{EO}$, 
$2021\,\mathrm{KN_2}$,
$2021\,\mathrm{KQ_2}$,
$2021\,\mathrm{TG_1}$,
$2021\,\mathrm{TL_{14}}$,
and $2021\,\mathrm{TQ_4}$.
Physical modeling of these candidates will be presented elsewhere.

\begin{longtable}{
  p{20mm}p{10mm}p{7mm}p{10mm}p{5mm}
  p{7mm}p{3mm}p{7mm}p{5mm}p{3mm}
  p{5mm}p{7mm}p{25mm}
  }
  \caption{Summary of observational results.$^{*}$.}
     \hline
    Object  &  \multicolumn{1}{c}{$H$}     &\multicolumn{1}{c}{$D$}   & \multicolumn{1}{r}{$N_{\mathrm{obs}}$} &\multicolumn{1}{r}{$n$} & \multicolumn{3}{c}{$P$}  & \multicolumn{3}{c}{$\Delta m$} &  a/b & Note\\
            &\multicolumn{1}{c}{(mag)}    &\multicolumn{1}{c}{(m)}   &                    &     & \multicolumn{3}{c}{(s)}  & \multicolumn{3}{c}{(mag)}    &      &      \\ 
  \hline
  \endfirsthead
  \hline
      Object  &  \multicolumn{1}{c}{$H$}     & \multicolumn{1}{c}{$D$}   & \multicolumn{1}{r}{$N_{\mathrm{obs}}$} & \multicolumn{1}{r}{$n$} & \multicolumn{3}{c}{$P$}  & \multicolumn{3}{c}{$\Delta m$} &  a/b & Note\\
              & \multicolumn{1}{c}{(mag)}    & \multicolumn{1}{c}{(m)}   &                    &     & \multicolumn{3}{c}{(s)}  & \multicolumn{3}{c}{(mag)}    &      &      \\ 
  \hline
  \endhead
  \hline
  \endfoot
  \label{tab:resneo1}
  
    \multicolumn{13}{l}{\footnotesize{
    $^{*}$ 
    $N_{\mathrm{obs}}$ is the number of frames.
    $n$ is the number of harmonics of the model curve. 
    $P$ is a rotational period.
    }}
    \\ \multicolumn{13}{l}{\footnotesize{
    $\Delta m$ is a light curve amplitude.
    $a/b$ is an axial ratio of the asteroid derived from $\Delta m$.
    starting time in UTC (Obs. Date)
    }}
    \\ \multicolumn{13}{l}{\footnotesize{
     and duration time of observation ($T$) 
    for each object are listed.
    $V$-band apparent magnitude ($V$), 
    }}
    \endlastfoot$2010\,\mathrm{WC_{9}}$ & \multicolumn{1}{c}{23.5}& \multicolumn{1}{c}{59}& \multicolumn{1}{r}{670}& \multicolumn{1}{r}{-}& \multicolumn{1}{r}{}&-& \multicolumn{1}{l}{\hspace{-3mm}}& \multicolumn{1}{r}{}&-& \multicolumn{1}{l}{\hspace{-3mm}}&- & known tumbler\\
$2011\,\mathrm{DW}$ & \multicolumn{1}{c}{22.9}& \multicolumn{1}{c}{79}& \multicolumn{1}{r}{2099}& \multicolumn{1}{r}{-}& \multicolumn{1}{r}{}&$>$1320& \multicolumn{1}{l}{\hspace{-3mm}}& \multicolumn{1}{r}{}&$>$1320& \multicolumn{1}{l}{\hspace{-3mm}}&- &  \\
$2017\,\mathrm{WJ_{16}}$ & \multicolumn{1}{c}{24.5}& \multicolumn{1}{c}{37}& \multicolumn{1}{r}{135}& \multicolumn{1}{r}{-}& \multicolumn{1}{r}{}&-& \multicolumn{1}{l}{\hspace{-3mm}}& \multicolumn{1}{r}{}&-& \multicolumn{1}{l}{\hspace{-3mm}}&- & tumbler\\
  &  &  & \multicolumn{1}{r}{2114}& \multicolumn{1}{r}{-}& \multicolumn{1}{r}{}&-& \multicolumn{1}{l}{\hspace{-3mm}}& \multicolumn{1}{r}{}&-& \multicolumn{1}{l}{\hspace{-3mm}}&- & tumbler\\
$2018\,\mathrm{LV_{3}}$ & \multicolumn{1}{c}{26.5}& \multicolumn{1}{c}{15}& \multicolumn{1}{r}{700}& \multicolumn{1}{r}{3}& \multicolumn{1}{r}{415.9}&$\pm$& \multicolumn{1}{l}{\hspace{-3mm}0.4}& \multicolumn{1}{r}{0.46}&$\pm$& \multicolumn{1}{l}{\hspace{-3mm}0.03}&$\geq$1.24 & 5 s exposure\\
$2018\,\mathrm{UD_{3}}$ & \multicolumn{1}{c}{26.2}& \multicolumn{1}{c}{17}& \multicolumn{1}{r}{3977}& \multicolumn{1}{r}{12}& \multicolumn{1}{r}{29.720}&$\pm$& \multicolumn{1}{l}{\hspace{-3mm}0.003}& \multicolumn{1}{r}{0.55}&$\pm$& \multicolumn{1}{l}{\hspace{-3mm}0.02}&$\geq$1.22 &  \\
$2019\,\mathrm{BE_{5}}$ & \multicolumn{1}{c}{25.1}& \multicolumn{1}{c}{28}& \multicolumn{1}{r}{1125}& \multicolumn{1}{r}{4}& \multicolumn{1}{r}{11.97902}&$\pm$& \multicolumn{1}{l}{\hspace{-3mm}0.00009}& \multicolumn{1}{r}{0.80}&$\pm$& \multicolumn{1}{l}{\hspace{-3mm}0.04}&$\geq$1.40 &  \\
$2020\,\mathrm{EO}$ & \multicolumn{1}{c}{25.9}& \multicolumn{1}{c}{20}& \multicolumn{1}{r}{1483}& \multicolumn{1}{r}{-}& \multicolumn{1}{r}{}&-& \multicolumn{1}{l}{\hspace{-3mm}}& \multicolumn{1}{r}{}&-& \multicolumn{1}{l}{\hspace{-3mm}}&- & tumbler\\
$2020\,\mathrm{FA_{2}}$ & \multicolumn{1}{c}{27.5}& \multicolumn{1}{c}{9}& \multicolumn{1}{r}{4930}& \multicolumn{1}{r}{6}& \multicolumn{1}{r}{150.67}&$\pm$& \multicolumn{1}{l}{\hspace{-3mm}0.02}& \multicolumn{1}{r}{0.26}&$\pm$& \multicolumn{1}{l}{\hspace{-3mm}0.01}&$\geq$1.15 &  \\
$2020\,\mathrm{FL_{2}}$ & \multicolumn{1}{c}{26.1}& \multicolumn{1}{c}{18}& \multicolumn{1}{r}{1557}& \multicolumn{1}{r}{9}& \multicolumn{1}{r}{325.1}&$\pm$& \multicolumn{1}{l}{\hspace{-3mm}0.8}& \multicolumn{1}{r}{0.102}&$\pm$& \multicolumn{1}{l}{\hspace{-3mm}0.005}&$\geq$1.07 &  \\
$2020\,\mathrm{GY_{1}}$ & \multicolumn{1}{c}{26.6}& \multicolumn{1}{c}{14}& \multicolumn{1}{r}{1474}& \multicolumn{1}{r}{6}& \multicolumn{1}{r}{303}&$\pm$& \multicolumn{1}{l}{\hspace{-3mm}3}& \multicolumn{1}{r}{0.17}&$\pm$& \multicolumn{1}{l}{\hspace{-3mm}0.02}&$\geq$1.10 &  \\
$2020\,\mathrm{HK_{3}}$ & \multicolumn{1}{c}{24.2}& \multicolumn{1}{c}{43}& \multicolumn{1}{r}{1476}& \multicolumn{1}{r}{-}& \multicolumn{1}{r}{}&$>$780& \multicolumn{1}{l}{\hspace{-3mm}}& \multicolumn{1}{r}{}&$>$780& \multicolumn{1}{l}{\hspace{-3mm}}&- &  \\
$2020\,\mathrm{HS_{7}}$ & \multicolumn{1}{c}{29.1}& \multicolumn{1}{c}{4}& \multicolumn{1}{r}{1365}& \multicolumn{1}{r}{1}& \multicolumn{1}{r}{2.9945}&$\pm$& \multicolumn{1}{l}{\hspace{-3mm}0.0002}& \multicolumn{1}{r}{0.069}&$\pm$& \multicolumn{1}{l}{\hspace{-3mm}0.006}&$\geq$1.04 &  \\
  &  &  & \multicolumn{1}{r}{896}& \multicolumn{1}{r}{2}& \multicolumn{1}{r}{2.9938}&$\pm$& \multicolumn{1}{l}{\hspace{-3mm}0.0002}& \multicolumn{1}{r}{0.075}&$\pm$& \multicolumn{1}{l}{\hspace{-3mm}0.006}&$\geq$1.04 &  \\
$2020\,\mathrm{HT_{7}}$ & \multicolumn{1}{c}{26.9}& \multicolumn{1}{c}{12}& \multicolumn{1}{r}{1411}& \multicolumn{1}{r}{6}& \multicolumn{1}{r}{45.8}&$\pm$& \multicolumn{1}{l}{\hspace{-3mm}0.01}& \multicolumn{1}{r}{0.38}&$\pm$& \multicolumn{1}{l}{\hspace{-3mm}0.02}&$\geq$1.17 &  \\
$2020\,\mathrm{HU_{3}}$ & \multicolumn{1}{c}{26.0}& \multicolumn{1}{c}{19}& \multicolumn{1}{r}{666}& \multicolumn{1}{r}{-}& \multicolumn{1}{r}{}&$>$360& \multicolumn{1}{l}{\hspace{-3mm}}& \multicolumn{1}{r}{}&$>$360& \multicolumn{1}{l}{\hspace{-3mm}}&- &  \\
$2020\,\mathrm{PW_{2}}$ & \multicolumn{1}{c}{28.8}& \multicolumn{1}{c}{5}& \multicolumn{1}{r}{1198}& \multicolumn{1}{r}{8}& \multicolumn{1}{r}{87.6}&$\pm$& \multicolumn{1}{l}{\hspace{-3mm}0.6}& \multicolumn{1}{r}{0.54}&$\pm$& \multicolumn{1}{l}{\hspace{-3mm}0.06}&$\geq$1.33 &  \\
$2020\,\mathrm{PY_{2}}$ & \multicolumn{1}{c}{26.5}& \multicolumn{1}{c}{15}& \multicolumn{1}{r}{1815}& \multicolumn{1}{r}{6}& \multicolumn{1}{r}{19.835}&$\pm$& \multicolumn{1}{l}{\hspace{-3mm}0.002}& \multicolumn{1}{r}{0.28}&$\pm$& \multicolumn{1}{l}{\hspace{-3mm}0.01}&$\geq$1.21 &  \\
$2020\,\mathrm{QW}$ & \multicolumn{1}{c}{25.3}& \multicolumn{1}{c}{26}& \multicolumn{1}{r}{1102}& \multicolumn{1}{r}{-}& \multicolumn{1}{r}{}&$>$1200& \multicolumn{1}{l}{\hspace{-3mm}}& \multicolumn{1}{r}{}&$>$1200& \multicolumn{1}{l}{\hspace{-3mm}}&- &  \\
$2020\,\mathrm{TD_{8}}$ & \multicolumn{1}{c}{26.9}& \multicolumn{1}{c}{12}& \multicolumn{1}{r}{434}& \multicolumn{1}{r}{5}& \multicolumn{1}{r}{29.53}&$\pm$& \multicolumn{1}{l}{\hspace{-3mm}0.01}& \multicolumn{1}{r}{1.19}&$\pm$& \multicolumn{1}{l}{\hspace{-3mm}0.04}&$\geq$1.56 &  \\
$2020\,\mathrm{TE_{6}}$ & \multicolumn{1}{c}{27.4}& \multicolumn{1}{c}{10}& \multicolumn{1}{r}{1537}& \multicolumn{1}{r}{-}& \multicolumn{1}{r}{}&-& \multicolumn{1}{l}{\hspace{-3mm}}& \multicolumn{1}{r}{}&-& \multicolumn{1}{l}{\hspace{-3mm}}&- & tumbler\\
$2020\,\mathrm{TS_{1}}$ & \multicolumn{1}{c}{29.2}& \multicolumn{1}{c}{4}& \multicolumn{1}{r}{793}& \multicolumn{1}{r}{-}& \multicolumn{1}{r}{}&$>$540& \multicolumn{1}{l}{\hspace{-3mm}}& \multicolumn{1}{r}{}&$>$540& \multicolumn{1}{l}{\hspace{-3mm}}&- &  \\
$2020\,\mathrm{UQ_{6}}$ & \multicolumn{1}{c}{22.7}& \multicolumn{1}{c}{86}& \multicolumn{1}{r}{1730}& \multicolumn{1}{r}{13}& \multicolumn{1}{r}{162.82}&$\pm$& \multicolumn{1}{l}{\hspace{-3mm}0.03}& \multicolumn{1}{r}{0.819}&$\pm$& \multicolumn{1}{l}{\hspace{-3mm}0.009}&$\geq$1.66 &  \\
$2020\,\mathrm{VF_{4}}$ & \multicolumn{1}{c}{26.6}& \multicolumn{1}{c}{14}& \multicolumn{1}{r}{1808}& \multicolumn{1}{r}{-}& \multicolumn{1}{r}{}&$>$1200& \multicolumn{1}{l}{\hspace{-3mm}}& \multicolumn{1}{r}{}&$>$1200& \multicolumn{1}{l}{\hspace{-3mm}}&- &  \\
$2020\,\mathrm{VH_{5}}$ & \multicolumn{1}{c}{29.2}& \multicolumn{1}{c}{4}& \multicolumn{1}{r}{2098}& \multicolumn{1}{r}{7}& \multicolumn{1}{r}{157.6}&$\pm$& \multicolumn{1}{l}{\hspace{-3mm}0.4}& \multicolumn{1}{r}{0.15}&$\pm$& \multicolumn{1}{l}{\hspace{-3mm}0.01}&$\geq$1.12 &  \\
$2020\,\mathrm{VJ_{1}}$ & \multicolumn{1}{c}{26.7}& \multicolumn{1}{c}{13}& \multicolumn{1}{r}{938}& \multicolumn{1}{r}{13}& \multicolumn{1}{r}{241}&$\pm$& \multicolumn{1}{l}{\hspace{-3mm}1}& \multicolumn{1}{r}{0.64}&$\pm$& \multicolumn{1}{l}{\hspace{-3mm}0.05}&$\geq$1.34 &  \\
$2020\,\mathrm{VR_{1}}$ & \multicolumn{1}{c}{28.9}& \multicolumn{1}{c}{5}& \multicolumn{1}{r}{677}& \multicolumn{1}{r}{-}& \multicolumn{1}{r}{}&$>$1200& \multicolumn{1}{l}{\hspace{-3mm}}& \multicolumn{1}{r}{}&$>$1200& \multicolumn{1}{l}{\hspace{-3mm}}&- &  \\
$2020\,\mathrm{VZ_{6}}$ & \multicolumn{1}{c}{25.0}& \multicolumn{1}{c}{30}& \multicolumn{1}{r}{1622}& \multicolumn{1}{r}{10}& \multicolumn{1}{r}{353.4}&$\pm$& \multicolumn{1}{l}{\hspace{-3mm}0.2}& \multicolumn{1}{r}{1.06}&$\pm$& \multicolumn{1}{l}{\hspace{-3mm}0.02}&$\geq$1.66 &  \\
$2020\,\mathrm{XH}$ & \multicolumn{1}{c}{24.6}& \multicolumn{1}{c}{36}& \multicolumn{1}{r}{208}& \multicolumn{1}{r}{-}& \multicolumn{1}{r}{}&$>$1020& \multicolumn{1}{l}{\hspace{-3mm}}& \multicolumn{1}{r}{}&$>$1020& \multicolumn{1}{l}{\hspace{-3mm}}&- &  \\
$2020\,\mathrm{XH_{1}}$ & \multicolumn{1}{c}{22.9}& \multicolumn{1}{c}{78}& \multicolumn{1}{r}{1912}& \multicolumn{1}{r}{-}& \multicolumn{1}{r}{}&$>$1200& \multicolumn{1}{l}{\hspace{-3mm}}& \multicolumn{1}{r}{}&$>$1200& \multicolumn{1}{l}{\hspace{-3mm}}&- &  \\
$2020\,\mathrm{XQ_{2}}$ & \multicolumn{1}{c}{22.8}& \multicolumn{1}{c}{83}& \multicolumn{1}{r}{244}& \multicolumn{1}{r}{-}& \multicolumn{1}{r}{}&$>$1200& \multicolumn{1}{l}{\hspace{-3mm}}& \multicolumn{1}{r}{}&$>$1200& \multicolumn{1}{l}{\hspace{-3mm}}&- &  \\
$2020\,\mathrm{XX_{3}}$ & \multicolumn{1}{c}{28.5}& \multicolumn{1}{c}{6}& \multicolumn{1}{r}{1664}& \multicolumn{1}{r}{13}& \multicolumn{1}{r}{136.22}&$\pm$& \multicolumn{1}{l}{\hspace{-3mm}0.05}& \multicolumn{1}{r}{0.98}&$\pm$& \multicolumn{1}{l}{\hspace{-3mm}0.02}&$\geq$1.52 &  \\
$2020\,\mathrm{XY_{4}}$ & \multicolumn{1}{c}{26.9}& \multicolumn{1}{c}{12}& \multicolumn{1}{r}{2020}& \multicolumn{1}{r}{3}& \multicolumn{1}{r}{324}&$\pm$& \multicolumn{1}{l}{\hspace{-3mm}6}& \multicolumn{1}{r}{0.15}&$\pm$& \multicolumn{1}{l}{\hspace{-3mm}0.01}&$\geq$1.06 &  \\
$2020\,\mathrm{YJ_{2}}$ & \multicolumn{1}{c}{27.4}& \multicolumn{1}{c}{10}& \multicolumn{1}{r}{182}& \multicolumn{1}{r}{-}& \multicolumn{1}{r}{}&$>$1200& \multicolumn{1}{l}{\hspace{-3mm}}& \multicolumn{1}{r}{}&$>$1200& \multicolumn{1}{l}{\hspace{-3mm}}&- &  \\
$2021\,\mathrm{AT_{5}}$ & \multicolumn{1}{c}{27.5}& \multicolumn{1}{c}{9}& \multicolumn{1}{r}{458}& \multicolumn{1}{r}{-}& \multicolumn{1}{r}{}&$>$600& \multicolumn{1}{l}{\hspace{-3mm}}& \multicolumn{1}{r}{}&$>$600& \multicolumn{1}{l}{\hspace{-3mm}}&- &  \\
$2021\,\mathrm{BC}$ & \multicolumn{1}{c}{24.3}& \multicolumn{1}{c}{41}& \multicolumn{1}{r}{1382}& \multicolumn{1}{r}{-}& \multicolumn{1}{r}{}&$>$1200& \multicolumn{1}{l}{\hspace{-3mm}}& \multicolumn{1}{r}{}&$>$1200& \multicolumn{1}{l}{\hspace{-3mm}}&- &  \\
$2021\,\mathrm{CA_{6}}$ & \multicolumn{1}{c}{28.5}& \multicolumn{1}{c}{6}& \multicolumn{1}{r}{2219}& \multicolumn{1}{r}{6}& \multicolumn{1}{r}{14.3159}&$\pm$& \multicolumn{1}{l}{\hspace{-3mm}0.0004}& \multicolumn{1}{r}{0.694}&$\pm$& \multicolumn{1}{l}{\hspace{-3mm}0.008}&$\geq$1.24 &  \\
$2021\,\mathrm{CC_{7}}$ & \multicolumn{1}{c}{29.8}& \multicolumn{1}{c}{3}& \multicolumn{1}{r}{1109}& \multicolumn{1}{r}{4}& \multicolumn{1}{r}{13.514}&$\pm$& \multicolumn{1}{l}{\hspace{-3mm}0.008}& \multicolumn{1}{r}{0.24}&$\pm$& \multicolumn{1}{l}{\hspace{-3mm}0.02}&$\geq$1.18 &  \\
$2021\,\mathrm{CG}$ & \multicolumn{1}{c}{26.1}& \multicolumn{1}{c}{18}& \multicolumn{1}{r}{1857}& \multicolumn{1}{r}{9}& \multicolumn{1}{r}{15.296}&$\pm$& \multicolumn{1}{l}{\hspace{-3mm}0.002}& \multicolumn{1}{r}{0.27}&$\pm$& \multicolumn{1}{l}{\hspace{-3mm}0.02}&$\geq$1.18 &  \\
$2021\,\mathrm{CO}$ & \multicolumn{1}{c}{25.3}& \multicolumn{1}{c}{26}& \multicolumn{1}{r}{1603}& \multicolumn{1}{r}{-}& \multicolumn{1}{r}{}&-& \multicolumn{1}{l}{\hspace{-3mm}}& \multicolumn{1}{r}{}&-& \multicolumn{1}{l}{\hspace{-3mm}}&- & known tumbler\\
$2021\,\mathrm{DW_{1}}$ & \multicolumn{1}{c}{25.2}& \multicolumn{1}{c}{27}& \multicolumn{1}{r}{146}& \multicolumn{1}{r}{4}& \multicolumn{1}{r}{23.8}&$\pm$& \multicolumn{1}{l}{\hspace{-3mm}0.2}& \multicolumn{1}{r}{0.68}&$\pm$& \multicolumn{1}{l}{\hspace{-3mm}0.08}&$\geq$1.29 &  \\
$2021\,\mathrm{EM_{4}}$ & \multicolumn{1}{c}{27.1}& \multicolumn{1}{c}{11}& \multicolumn{1}{r}{1438}& \multicolumn{1}{r}{8}& \multicolumn{1}{r}{99.5}&$\pm$& \multicolumn{1}{l}{\hspace{-3mm}0.4}& \multicolumn{1}{r}{0.26}&$\pm$& \multicolumn{1}{l}{\hspace{-3mm}0.03}&$\geq$1.14 &  \\
$2021\,\mathrm{EQ_{3}}$ & \multicolumn{1}{c}{26.1}& \multicolumn{1}{c}{18}& \multicolumn{1}{r}{1376}& \multicolumn{1}{r}{11}& \multicolumn{1}{r}{119.41}&$\pm$& \multicolumn{1}{l}{\hspace{-3mm}0.02}& \multicolumn{1}{r}{0.71}&$\pm$& \multicolumn{1}{l}{\hspace{-3mm}0.02}&$\geq$1.30 &  \\
$2021\,\mathrm{ET_{4}}$ & \multicolumn{1}{c}{23.9}& \multicolumn{1}{c}{48}& \multicolumn{1}{r}{1703}& \multicolumn{1}{r}{4}& \multicolumn{1}{r}{87.8}&$\pm$& \multicolumn{1}{l}{\hspace{-3mm}0.2}& \multicolumn{1}{r}{0.21}&$\pm$& \multicolumn{1}{l}{\hspace{-3mm}0.02}&$\geq$1.09 &  \\
$2021\,\mathrm{EX_{1}}$ & \multicolumn{1}{c}{24.9}& \multicolumn{1}{c}{32}& \multicolumn{1}{r}{2227}& \multicolumn{1}{r}{4}& \multicolumn{1}{r}{410}&$\pm$& \multicolumn{1}{l}{\hspace{-3mm}1}& \multicolumn{1}{r}{0.222}&$\pm$& \multicolumn{1}{l}{\hspace{-3mm}0.009}&$\geq$1.13 &  \\
$2021\,\mathrm{FH}$ & \multicolumn{1}{c}{26.7}& \multicolumn{1}{c}{13}& \multicolumn{1}{r}{1632}& \multicolumn{1}{r}{3}& \multicolumn{1}{r}{63.5}&$\pm$& \multicolumn{1}{l}{\hspace{-3mm}0.6}& \multicolumn{1}{r}{0.16}&$\pm$& \multicolumn{1}{l}{\hspace{-3mm}0.02}&$\geq$1.09 &  \\
$2021\,\mathrm{GD_{5}}$ & \multicolumn{1}{c}{27.1}& \multicolumn{1}{c}{11}& \multicolumn{1}{r}{2240}& \multicolumn{1}{r}{-}& \multicolumn{1}{r}{}&$>$1200& \multicolumn{1}{l}{\hspace{-3mm}}& \multicolumn{1}{r}{}&$>$1200& \multicolumn{1}{l}{\hspace{-3mm}}&- &  \\
$2021\,\mathrm{GQ_{10}}$ & \multicolumn{1}{c}{26.6}& \multicolumn{1}{c}{14}& \multicolumn{1}{r}{1167}& \multicolumn{1}{r}{3}& \multicolumn{1}{r}{19.308}&$\pm$& \multicolumn{1}{l}{\hspace{-3mm}0.003}& \multicolumn{1}{r}{0.192}&$\pm$& \multicolumn{1}{l}{\hspace{-3mm}0.007}&$\geq$1.06 &  \\
$2021\,\mathrm{GT_{3}}$ & \multicolumn{1}{c}{26.4}& \multicolumn{1}{c}{16}& \multicolumn{1}{r}{2013}& \multicolumn{1}{r}{10}& \multicolumn{1}{r}{155.1}&$\pm$& \multicolumn{1}{l}{\hspace{-3mm}0.2}& \multicolumn{1}{r}{0.149}&$\pm$& \multicolumn{1}{l}{\hspace{-3mm}0.009}&$\geq$1.11 &  \\
$2021\,\mathrm{JB_{6}}$ & \multicolumn{1}{c}{28.8}& \multicolumn{1}{c}{5}& \multicolumn{1}{r}{2212}& \multicolumn{1}{r}{2}& \multicolumn{1}{r}{65.64}&$\pm$& \multicolumn{1}{l}{\hspace{-3mm}0.02}& \multicolumn{1}{r}{0.55}&$\pm$& \multicolumn{1}{l}{\hspace{-3mm}0.01}&$\geq$1.23 &  \\
$2021\,\mathrm{KN_{2}}$ & \multicolumn{1}{c}{28.6}& \multicolumn{1}{c}{6}& \multicolumn{1}{r}{1244}& \multicolumn{1}{r}{-}& \multicolumn{1}{r}{}&-& \multicolumn{1}{l}{\hspace{-3mm}}& \multicolumn{1}{r}{}&-& \multicolumn{1}{l}{\hspace{-3mm}}&- & known tumbler\\
$2021\,\mathrm{KQ_{2}}$ & \multicolumn{1}{c}{29.9}& \multicolumn{1}{c}{3}& \multicolumn{1}{r}{1829}& \multicolumn{1}{r}{-}& \multicolumn{1}{r}{}&-& \multicolumn{1}{l}{\hspace{-3mm}}& \multicolumn{1}{r}{}&-& \multicolumn{1}{l}{\hspace{-3mm}}&- & tumbler\\
$2021\,\mathrm{RB_{1}}$ & \multicolumn{1}{c}{24.1}& \multicolumn{1}{c}{46}& \multicolumn{1}{r}{1931}& \multicolumn{1}{r}{-}& \multicolumn{1}{r}{}&$>$1200& \multicolumn{1}{l}{\hspace{-3mm}}& \multicolumn{1}{r}{}&$>$1200& \multicolumn{1}{l}{\hspace{-3mm}}&- &  \\
$2021\,\mathrm{RX_{5}}$ & \multicolumn{1}{c}{23.7}& \multicolumn{1}{c}{54}& \multicolumn{1}{r}{528}& \multicolumn{1}{r}{-}& \multicolumn{1}{r}{}&$>$420& \multicolumn{1}{l}{\hspace{-3mm}}& \multicolumn{1}{r}{}&$>$420& \multicolumn{1}{l}{\hspace{-3mm}}&- &  \\
$2021\,\mathrm{TG_{1}}$ & \multicolumn{1}{c}{28.2}& \multicolumn{1}{c}{7}& \multicolumn{1}{r}{780}& \multicolumn{1}{r}{-}& \multicolumn{1}{r}{}&-& \multicolumn{1}{l}{\hspace{-3mm}}& \multicolumn{1}{r}{}&-& \multicolumn{1}{l}{\hspace{-3mm}}&- & tumbler\\
$2021\,\mathrm{TL_{14}}$ & \multicolumn{1}{c}{26.9}& \multicolumn{1}{c}{12}& \multicolumn{1}{r}{1626}& \multicolumn{1}{r}{-}& \multicolumn{1}{r}{}&-& \multicolumn{1}{l}{\hspace{-3mm}}& \multicolumn{1}{r}{}&-& \multicolumn{1}{l}{\hspace{-3mm}}&- & tumbler\\
$2021\,\mathrm{TQ_{3}}$ & \multicolumn{1}{c}{27.1}& \multicolumn{1}{c}{11}& \multicolumn{1}{r}{1897}& \multicolumn{1}{r}{-}& \multicolumn{1}{r}{}&$>$1200& \multicolumn{1}{l}{\hspace{-3mm}}& \multicolumn{1}{r}{}&$>$1200& \multicolumn{1}{l}{\hspace{-3mm}}&- &  \\
$2021\,\mathrm{TQ_{4}}$ & \multicolumn{1}{c}{29.9}& \multicolumn{1}{c}{3}& \multicolumn{1}{r}{424}& \multicolumn{1}{r}{-}& \multicolumn{1}{r}{}&-& \multicolumn{1}{l}{\hspace{-3mm}}& \multicolumn{1}{r}{}&-& \multicolumn{1}{l}{\hspace{-3mm}}&- & tumbler\\
$2021\,\mathrm{TY_{14}}$ & \multicolumn{1}{c}{27.2}& \multicolumn{1}{c}{11}& \multicolumn{1}{r}{2051}& \multicolumn{1}{r}{4}& \multicolumn{1}{r}{15.292}&$\pm$& \multicolumn{1}{l}{\hspace{-3mm}0.002}& \multicolumn{1}{r}{0.61}&$\pm$& \multicolumn{1}{l}{\hspace{-3mm}0.02}&$\geq$1.40 &  \\
$2021\,\mathrm{UF_{12}}$ & \multicolumn{1}{c}{29.3}& \multicolumn{1}{c}{4}& \multicolumn{1}{r}{423}& \multicolumn{1}{r}{1}& \multicolumn{1}{r}{14.86}&$\pm$& \multicolumn{1}{l}{\hspace{-3mm}0.004}& \multicolumn{1}{r}{0.51}&$\pm$& \multicolumn{1}{l}{\hspace{-3mm}0.02}&$\geq$1.39 &  \\
TMG0042 & \multicolumn{1}{c}{28.5}& \multicolumn{1}{c}{6}& \multicolumn{1}{r}{1937}& \multicolumn{1}{r}{20}& \multicolumn{1}{r}{314.4}&$\pm$& \multicolumn{1}{l}{\hspace{-3mm}0.3}& \multicolumn{1}{r}{1.00}&$\pm$& \multicolumn{1}{l}{\hspace{-3mm}0.04}&$\geq$2.51 &  \\
TMG0049 & \multicolumn{1}{c}{30.0}& \multicolumn{1}{c}{3}& \multicolumn{1}{r}{1538}& \multicolumn{1}{r}{-}& \multicolumn{1}{r}{}&$>$1080& \multicolumn{1}{l}{\hspace{-3mm}}& \multicolumn{1}{r}{}&$>$1080& \multicolumn{1}{l}{\hspace{-3mm}}&- &  \\

    \hline
    \end{longtable}

\subsection{Light curves and periodograms}
\begin{figure*}
  \includegraphics[width=170mm]{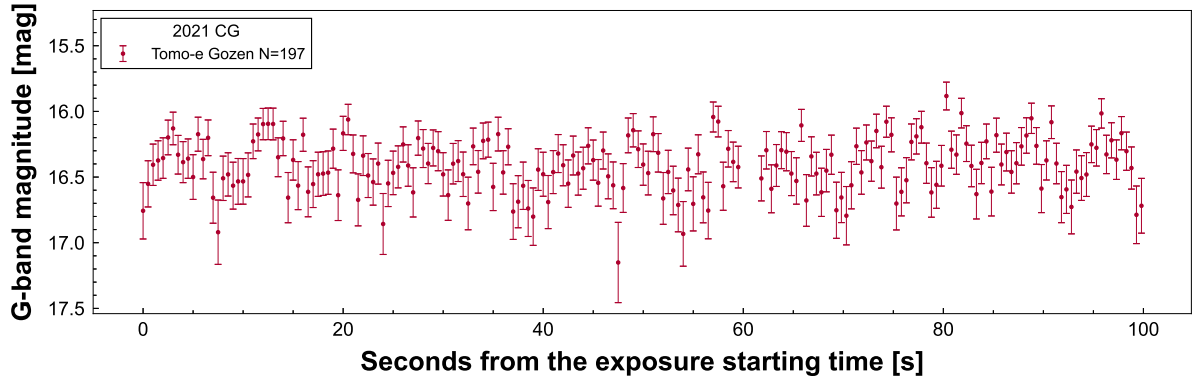}
  \caption{%
      Light curve of $2021\,\mathrm{CG}$.
    The first 100\,s part of the whole 20\,minutes light curve are plotted.
    Bars indicate the 1 sigma uncertainties (see text for details).
    }\label{fig:lc2021CG}
\end{figure*}

\begin{figure}
    \includegraphics[width=80mm]{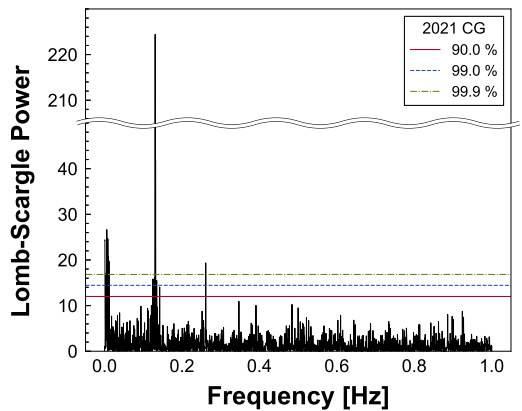}
    \caption{%
        Lomb-Scargle periodogram of $2021\,\mathrm{CG}$. The number of harmonics is unity.
      Solid, dashed, and dot-dashed horizontal lines show
      90.0, 99.0, and 99.9\% confidence levels, respectively.
    }\label{fig:LS2021CG}
\end{figure}

\begin{figure}
    \includegraphics[width=80mm]{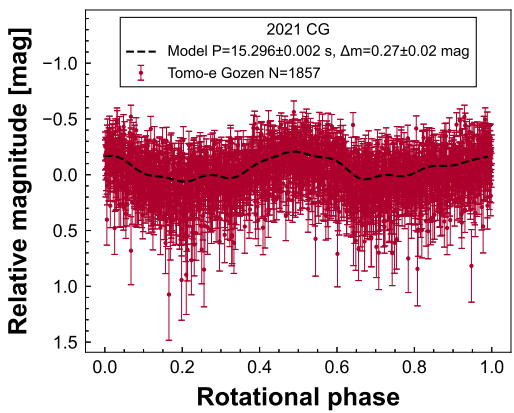}
    \caption{%
        Phased light curve of $2021\,\mathrm{CG}$.
      A model curve with a period of 15.296\,s and a light curve amplitude of 0.27\,mag
      is shown by a dashed line.
      Photometric errors are the same as in figure \ref{fig:lc2021CG}. 
      }\label{fig:plc2021CG}
\end{figure}

As an example, we presented the light curve and periodogram of $2021\,\mathrm{CG}$ 
in figures \ref{fig:lc2021CG} and \ref{fig:LS2021CG}, respectively.
The rotational period and the light curve amplitude of $2021\,\mathrm{CG}$ 
were estimated to be $15.296\pm0.002\,\mathrm{s}$ and $0.27\pm0.02$\,mag, respectively,
using the Monte Carlo method as shown in figure \ref{fig:MC2021CG}.
The light curve folded by the rotational period (hereinafter referred to as phased light curve) 
is shown in figure \ref{fig:plc2021CG}. 
Thanks to the video observations at 2\,fps, 
we can estimate such a short period of rotation with high reliability.
The light curves, periodograms of 60 NEOs,
and phased light curves of NEOs whose rotational periods were estimated 
are attached in figures (\ref{fig:rotPest})--(\ref{fig:rotPest_tumbler}), see the Appendix.

\subsection{D-P relation}
\begin{figure*}
  \includegraphics[width=160mm]{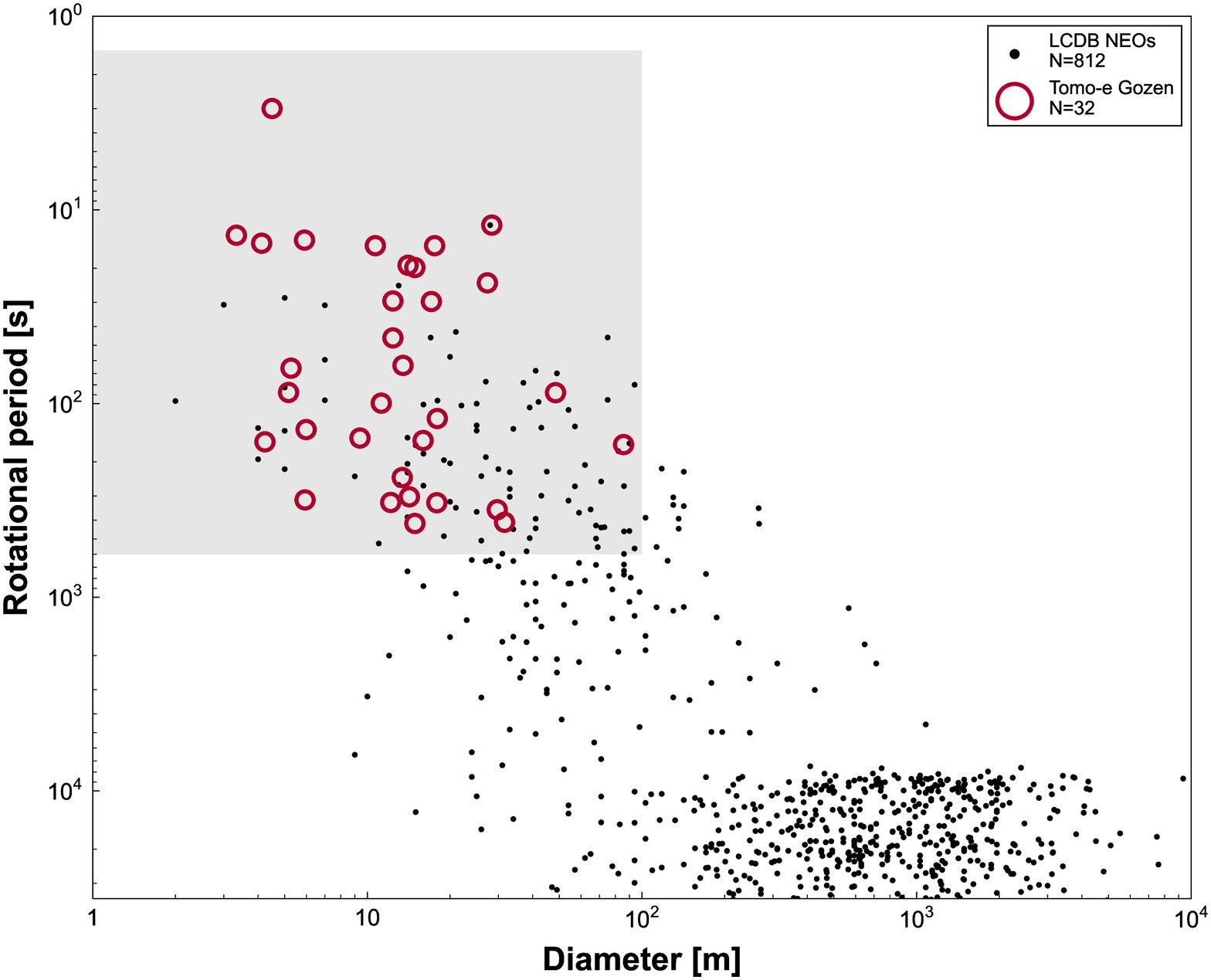}
  \caption{%
    D-P relations of the Tomo-e NEOs (open circles) and the NEOs in LCDB (filled circles).
    The range of detectable rotational period of our targets ($D \leq 100$\,m), 1.5\,s to 10\,min, in typical observations at 2\,fps for 20\,min
    with Tomo-e Gozen is shown as a gray shaded area.
    }\label{fig:D-Presult}
\end{figure*}

The D-P relation of the Tomo-e NEOs and the NEOs in LCDB is shown in figure \ref{fig:D-Presult}.
The Tomo-e NEOs are distributed in a range of 3 to 100\,m in diameter and 
3 to 420\,s in rotational period.
We found 13 NEOs with rotational periods less than 60\,s.

\begin{figure}
  \includegraphics[width=80mm]{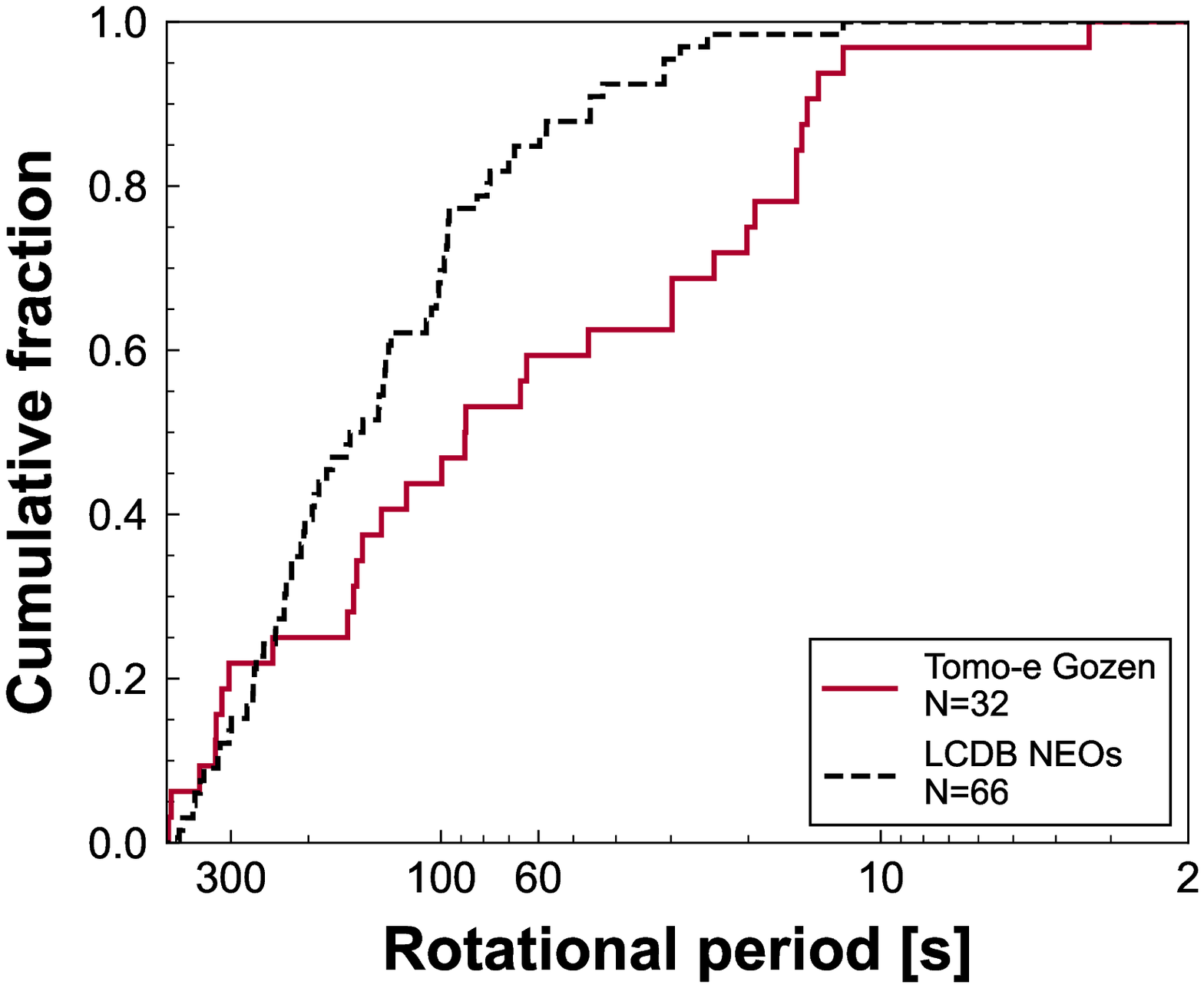}
  \caption{%
    Cumulative histograms of rotational periods of 
    the Tomo-e Gozen NEOs (solid line) and the NEOs in LCDB (dashed line)
    with absolute magnitudes larger than 22.5 and rotational periods shorter than
    420\,s.
    }\label{fig:KS}
\end{figure}

We create cumulative histograms of rotational periods of the Tomo-e NEOs and
the NEOs in LCDB (figure \ref{fig:KS}).
The D-P relation of the Tomo-e NEOs looks different from that of the NEOs in LCDB.
We performed the Kolmogorov-Smirnov(KS) test
to check the null hypothesis that the two D-P relations are the same.
We chose the NEOs satisfying the criteria that
the absolute magnitude is larger than 22.5,
the rotational period is shorter than 420\,s corresponding to
the longest rotational period of the Tomo-e NEOs.
The NEOs whose quality code is \texttt{3} or \texttt{3-} are used as for the NEOs in LCDB.
The deduced KS statistics and the $p$-value are 0.330 and 0.013, respectively.
This tentatively implies that rotational periods of some fast rotators 
have not been able to be estimated due to 
long exposure times and other factors in the previous studies.

\subsection{Axial ratios}
We defined a light curve amplitude $\Delta m$ by
a difference between maximum and minimum values of the model curve.
We assumed the asteroid is a triaxial ellipsoid with axial lengths of $a$, $b$, and $c$ ($a > b > c$)
and the aspect angle of $90^{\circ}$.
A lower limit of axial ratio $a/b$ is estimated as follows:
\begin{equation}
    \frac{a}{b} \geq
    10^{0.4\Delta m(\alpha)/(1+s\alpha)},
    \label{eq:ab}
\end{equation}
where $\Delta m(\alpha)$ is the light curve amplitude at a phase angle of $\alpha$
and $s$ is a slope depending on the taxonomic type of the asteroid \citep{Bowell1989}.
We assumed that $s$ is 0.030, a typical value of S-type asteroids \citep{Zappala1990}.

\begin{figure*}
  \begin{minipage}{0.48\linewidth}
    \includegraphics[width=80mm]{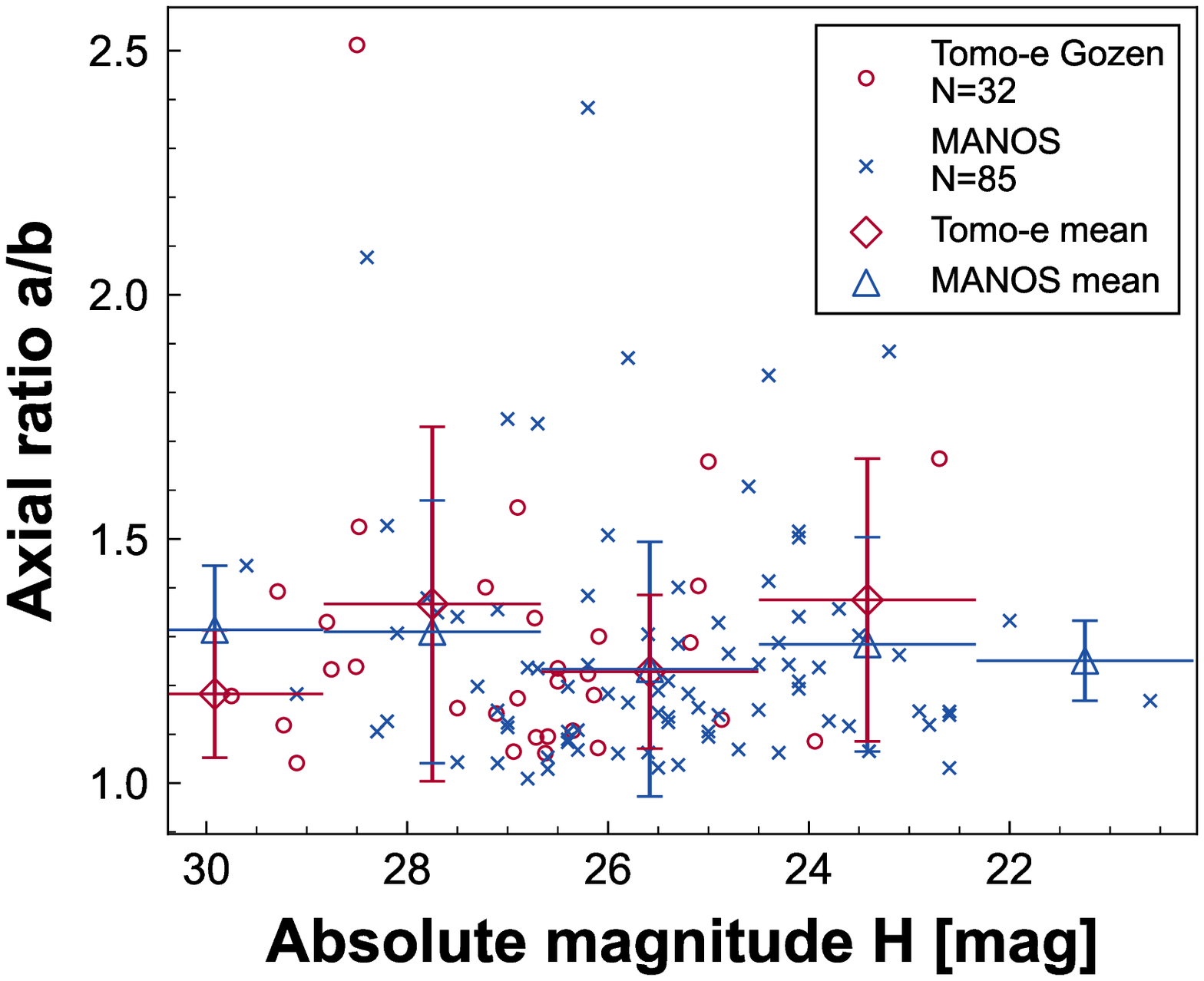}
    \caption{%
      Absolute magnitude versus lower limits of axial ratios of 
      the Tomo-e (open circles) and the MANOS NEOs (crosses) with rotational 
      periods shorter than 600\,s. 
      The mean value in each range is presented by a diamond and a triangle
      for the Tomo-e NEOs and the MANOS NEOs, respectively.
      Vertical bars indicate standard deviations.
      }\label{fig:H_ab}
  \end{minipage}
  \begin{minipage}{0.48\linewidth}
    \includegraphics[width=80mm]{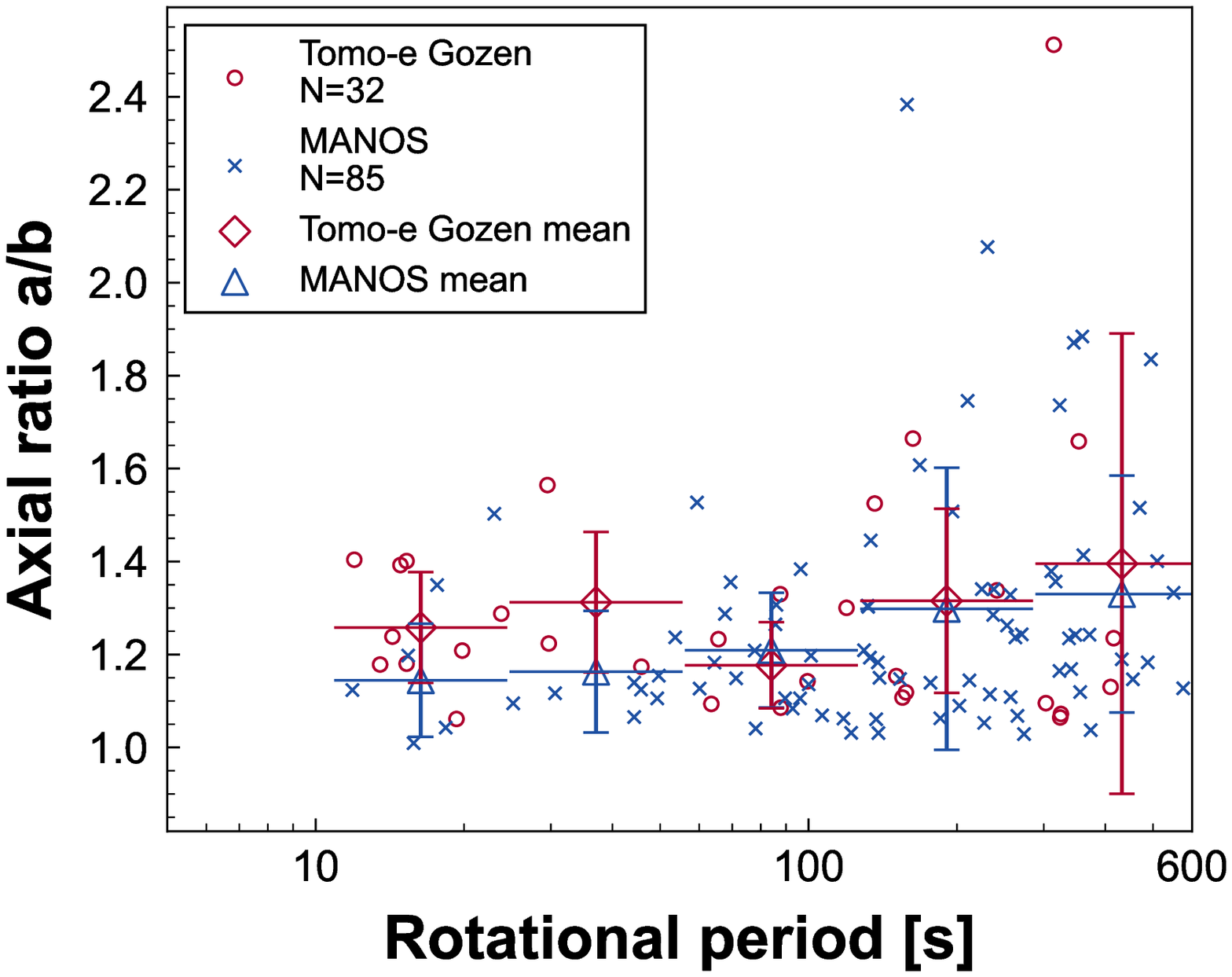}
    \caption{Rotational periods versus lower limits of axial ratios 
      of the Tomo-e (open circles) and the MANOS NEOs (crosses) with rotational
      periods shorter than 600\,s. 
      The mean value in each range is presented by a diamond and a triangle
      for the Tomo-e NEOs and the MANOS NEOs, respectively.
      Vertical bars indicate standard deviations.
      }\label{fig:P_ab}
  \end{minipage}
\end{figure*}

A relation between the absolute magnitudes $H$
and the lower limits of axial ratios $a/b$ of the Tomo-e NEOs and the MANOS NEOs 
with rotational periods shorter than 600\,s 
are shown in figure \ref{fig:H_ab}.
A relation between the rotational period $P$ and $a/b$
of the Tomo-e NEOs and the MANOS NEOs 
with $P \leq 600\,\mathrm{s}$ are shown in figure \ref{fig:P_ab}.
The mean of $a/b$ for each range of $H$ and $P$ is also presented.
The range is determined based on the Sturges' rule.
No strong correlation is seen in both figures \ref{fig:H_ab} and \ref{fig:P_ab}.
The present results are consistent with
Hatch \& Wiegert\,(\yearcite{Hatch2015}) and \citet{Thirouin2016}.

The difference of mean axial ratios 
between the Tomo-e NEOs ($\sim$ 1.29) and the MANOS NEOs ($\sim$ 1.27) is about 0.02.
We performed a bootstrap test to check the null hypothesis that the mean axial ratios of two samples are the same.
We generated 10000 differences of the mean axial ratios by resampling the Tomo-e and MANOS NEOs.
The 95\,\% confidence interval is from -0.08 to 0.14.
Thus, the null hypothesis is not rejected at 5\,\% significance level.

\begin{figure}
  \includegraphics[width=80mm]{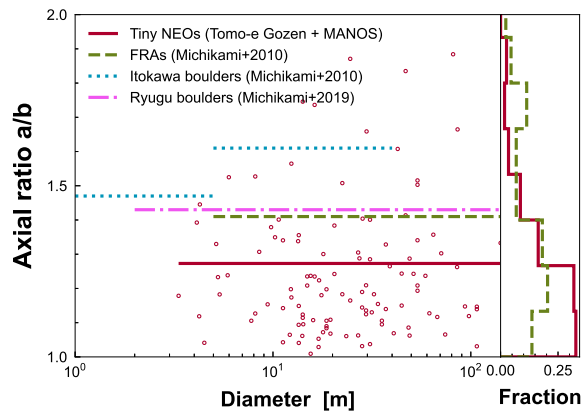}
  \caption{%
  Diameter versus lower limits of axial ratios of tiny NEOs, fast-rotating asteroids (FRAs),
    and apparent axis ratios of boulders on Itokawa and Ryugu.
    Mean values of tiny NEOs and FRAs are presented as solid and dashed lines, respectively.
    Typical values of boulders on Itokawa and Ryugu are presented as dotted and dash-dot lines, respectively.
    Fractions of each range of axial ratio for tiny NEOs and FRAs 
    are shown as histograms in right panel.
    NEOs with axial ratios larger than 2.0 are not shown in figures 
    but used in calculations.
    }\label{fig:comp_ba}
\end{figure}

Figure \ref{fig:comp_ba} shows
measured $a/b$ of various sources:
the average of the sum of the Tomo-e NEOs and the MANOS NEOs,
the average of fast-rotating asteroids (FRAs) with diameters less than 200\,m and 
a rotational period less than 1\,hour \citep{Michikami2010},
and the averages of boulders on the surfaces of asteroids Itokawa \citep{Michikami2010} and Ryugu \citep{Michikami2019}.

\citet{Michikami2010} mention that the lower limits of $a/b$ of FRAs and 
the $a/b$ of boulders are similar to those of laboratory experiments ($\sim$ 1.4),
although the aspect angles of asteroids are unknown.
The lower limits of $a/b$ of asteroids are lower in the case of recent observational results 
such as Tomo-e Gozen ($\sim$ 1.29) and MANOS ($\sim$ 1.27).
It is important not only to increase the number of light curve observations,
but also to determine pole directions to discuss a relation with fragments of laboratory experiments and 
boulders (e.g., \cite{Kwiatkowski2021}).

\section{Discussion}
\subsection{Detectable rotational period}
Sparse sampling and finite exposure time may lead to underestimating the light
curve amplitude and misidentifying periodicity
(Pravec \& Harris\,\yearcite{Pravec2000}; \cite{Thirouin2018, Birtwhistle2021c}).
We examine detection limits in rotational periods ($P_\mathrm{det}$) in our observations 
to verify the deficiency of asteroids rotating faster than 10\,s. 
We simulate light curves if an asteroid was rotating faster than it is, 
and then the light curves are analyzed in the same manner. 
We selected 8 NEOs such that their rotational periods are short ($P \leq 60\,\mathrm{s}$),
and the durations of the observations are typical ($10 \leq T \leq 30\,\mathrm{min}$).
We excluded $2020\,\mathrm{HS_7}$ in this examination 
since the exposure time is not sufficiently shorter than the rotational period and
the observed light curve can be underestimated \citep{Birtwhistle2021c}.
Their model light curves are used as templates. 
The rotational period of a hypothetical asteroid $P_\mathrm{pseudo}$ is set to 
$P_\mathrm{pseudo} = P/2, P/3, P/4,\,...$, where $P$ is the original rotational period in section \ref{sec3}.
Then, the hypothetical asteroid is virtually observed to generate a pseudo light curve. 
The number of measurements ($N_\mathrm{obs}$) and the timestamps are the same as the actual observation. 
The pseudo light curve is perturbed to match the noise level with the original observation.
The criterion of the periodicity identification is the same as in subsubsection \ref{periana}.

The results of the periodic analysis are summarized in table \ref{tab:detP}. 
The detectable rotational periods are less than 2\,s for all the 8 asteroids. 
Although our observations are unevenly sampled because of intervals between frames, 
large fractions of the data are evenly sampled at 2\,fps. 
The periodograms of the pseudo light curves can be affected by aliases 
when the frequency gets closer to the Nyquist limit $f_\mathrm{Ny} = 2/2 = 1\,\mathrm{Hz}$. 
However, the peaks by the aliases become weaker than the genuine peaks 
due to the uneven sampling. 
Thus, it is natural that we detect shorter rotational periods than the Nyquist limit 
($P = 2\,\mathrm{s}$ assuming typical double-peak light curves). 
We conservatively set the detectable rotational period to 1.5\,s in our systematic 
20\,min video observations at 2\,fps. 
Therefore, it is inevitable that there is only one fast rotator whose 
rotational period is shorter than 10\,s in our 60 NEOs.

\begin{table}
  \tbl{Periodic analysis results of pseudo light curves.\footnotemark[$*$]}{%
  \begin{tabular}{lcr@{\hspace{1mm}}c@{\hspace{1mm}}l}
    \hline
    Object               & $N_{\mathrm{det}}$ & \multicolumn{3}{c}{$P_{\mathrm{det}}$}             \\
                       &                    & \multicolumn{3}{c}{(s)}                            \\ 
    \hline
    pseudo $2020\,\mathrm{HT_7}$    &  1411   & 1.30906&$\pm$&0.0002                              \\ 
    pseudo $2020\,\mathrm{PY_2}$    &  1815   & 1.32235&$\pm$&0.00002                             \\ 
    pseudo $2020\,\mathrm{TD_8}$    &  434    & 1.2302&$\pm$&0.0001                               \\ 
    pseudo $2021\,\mathrm{CA_6}$    &  2219   & 1.31436&$\pm$&0.000007                            \\ 
    pseudo $2021\,\mathrm{CC_7}$    &  1109   & 1.501&$\pm$&0.001                                 \\ 
    pseudo $2021\,\mathrm{CG}$      &  1857   & 1.3906&$\pm$&0.0002                               \\ 
    pseudo $2021\,\mathrm{GQ_{10}}$ &  1167   & 1.20683&$\pm$&0.00004                               \\ 
    pseudo $2021\,\mathrm{TY_{14}}$ &  2051   & 1.27431&$\pm$&0.00006                              \\ \hline
  \end{tabular}
  }\label{tab:detP}
  \begin{tabnote}
    \footnotemark[$*$] 
    $P_{\mathrm{det}}$ is the detectable rotational period of the object with the 
    same observational conditions in this paper.
  \end{tabnote}
\end{table}

\subsection{Deficiency of fast rotators}
We found no NEOs with rotational periods shorter than 10\,s other than $2020\,\mathrm{HS_7}$.
The distribution of the Tomo-e NEOs in the D-P relation is truncated
around 10\,s in the rotational period as shown in figures \ref{fig:D-Presult}.
To interpret this flat-top distribution,
we consider the evolution of rotational periods of the NEOs.

Since smaller asteroids experience stronger Yarkovsky effect and their semi-major axes are changed,
parts of them drift to the resonances
with giant planets in the main belt in a short time scale ($\sim$ a few Myr) and
then are scattered into the near-Earth region \citep{Bottke2006}.
The orbits of the scattered asteroids evolve to those of NEOs over a few\,Myr \citep{Gladman1997}.
Therefore, typical NEOs are considered a few to 10\,Myr old.
This timescale (hereinafter referred to as NEO age) is
consistent with a typical cosmic ray exposure age of meteorites \citep{Eugster2006}.

Since YORP gradually changes the rotational states of NEOs during the orbital evolution,
the distribution of the rotational periods reflects the NEO age.
Although YORP also decelerates the rotation, here we consider only the acceleration.
The decelerated tiny asteroids shortly enter tumbling states once spinning down starts
(\cite{Vokrouhlicky2007, Breiter2011})
and it is difficult to predict their evolution accurately.
In this study, we estimate reachable rotational periods of NEOs by the YORP acceleration.
For the sake of the simplicity, we do not take into account the time evolution of
the orbital elements, resulting in a constant acceleration.

We use two assumptions as follows.
A tiny asteroid is a fragment of a collisional event and
its initial rotational period, $P_{\mathrm{init}}$,
is given by an extrapolation of the diameter and rotational period relation for
mm-sized fragments in a collisional experiment \citep{Kadono2009}:
\begin{equation}
    P_{\mathrm{init}} = 10 \left(\frac{D}{\mathrm{1\,m}}\right) \,\,\mathrm{s}.
    \label{eq:kadonoline}
\end{equation}
The YORP acceleration follows a scaling law and is derived from
the YORP strength of the near-Earth object Bennu \citep{Vokrouhlicky2004, Hergenrother2019}:
\begin{eqnarray}
    & \frac{d\omega}{dt} = 
    8.5 \times 10^{-18}
    \nonumber 
    \\
    & \times 
    \left(\frac{a_{\mathrm{Bennu}}^2 \sqrt{1-e_{\mathrm{Bennu}}^2}}
    {a_{\mathrm{ast}}^2 \sqrt{1-e_{\mathrm{ast}}^2}}\right)
    \left(\frac{D_{\mathrm{Bennu}}}{D}\right)^2 
\,\,\mathrm{rad\,s^{-2}},
    \label{eq:dw_YORP}
\end{eqnarray}
where $\omega$ is the angular velocity of the asteroid,
$D_{\mathrm{Bennu}}$
is the diameter of Bennu,
$a_{\mathrm{Bennu}}$ and $a_{\mathrm{ast}}$ are the semi-major axes
of Bennu and the asteroid,
and
$e_{\mathrm{Bennu}}$ and $e_{\mathrm{ast}}$ are the orbital eccentricities
of Bennu and the asteroid.
We adopt that $D_{\mathrm{Bennu}}$ is 482\,m,
$a_{\mathrm{Bennu}}$ is 1.126\,au,
and $e_{\mathrm{Bennu}}$ is 0.204 (JPL Small-Body Database
\footnote{\url{https://ssd.jpl.nasa.gov/tools/sbdb_lookup.html#/}(accessed 2021-12-20)}).
We set $a_{\mathrm{ast}}$ to 2\,au and $e_{\mathrm{ast}}$ to zero since most NEOs
have come from the inner main belt \citep{Granvik2018}.

We assume a linear acceleration of a rotational period by YORP 
and obtain the $\omega$ in time $t$ as follows:
\begin{equation}
    \omega = \frac{d\omega}{dt} t + \omega_0 \,\,\mathrm{rad\,s^{-1}},
\label{eq:omega}
\end{equation}
where $\omega_0$ is the initial angular velocity of the asteroid.
We calculate the NEO age, $\tau_\mathrm{YORP}$, as follows by solving the 
equation (\ref{eq:omega}) for $t$ with equations 
(\ref{eq:kadonoline}) and (\ref{eq:dw_YORP}):

\begin{eqnarray}
  & \tau_{\mathrm{YORP}} 
  = 3.7 \times 10^{3}
  \left(\frac{a_{\mathrm{ast}}^2 \sqrt{1-e_{\mathrm{ast}}^2}}
    {a_{\mathrm{Bennu}}^2 \sqrt{1-e_{\mathrm{Bennu}}^2}}\right)
    \\
  & \times 
    \left(\frac{D}{D_{\mathrm{Bennu}}}\right)^2 
    \left(\frac{1}{P}-\frac{1}{10D}\right)
    \,\,\mathrm{Myr}.
  \label{tau_YORP}
\end{eqnarray}

\begin{figure*}
  \includegraphics[width=160mm]{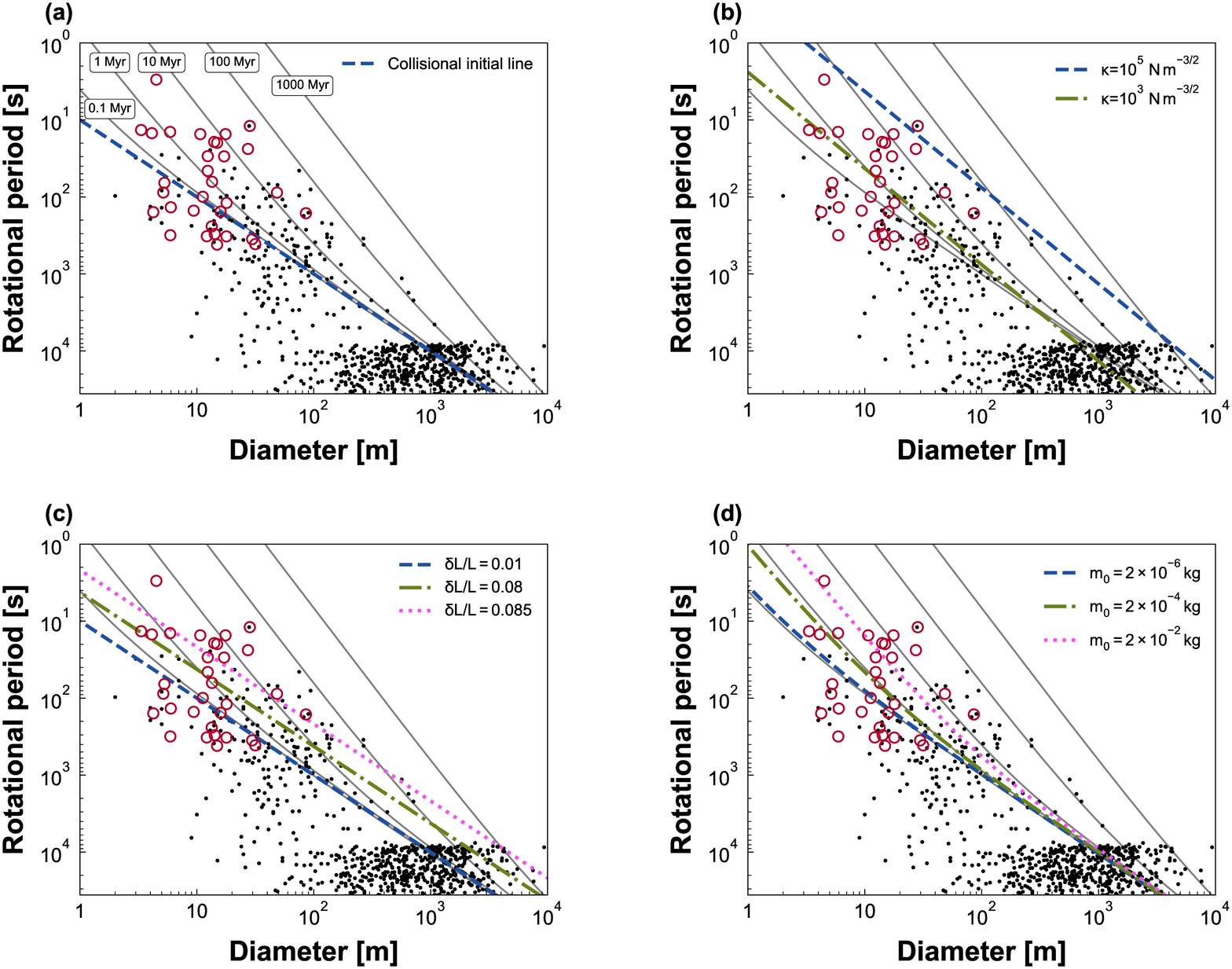}
    \caption{%
      D-P relations of NEOs with the isochrones (solid lines).
      The Tomo-e NEOs and the NEOs in LCDB are presented with
      open and filled circles, respectively.
      (a) 
      D-P relation with the collisional initial line (dashed line). 
      (b) 
      D-P relation with lines of critical rotational periods
      when asteroids have tensile strength of
      a typical meteorite (dashed line) and weak material (dot-dashed line).
      (c) 
      D-P relation with lower limits by angular momentum transfers due to meteoroid impacts 
      with 
      $\delta L/L =$ 0.01 (dashed-line), 0.08 (dot-dashed line), 
      and 0.085 (dotted line). 
      (d) 
      D-P relation with lower limits by cratering due to meteoroid impacts
      with 
      $m_0$ of $2\times10^{-6}$ kg (dashed-line), 
      $2\times10^{-4}$\,kg (dot-dashed line), 
      and $2\times10^{-2}$\,kg (dotted line).
      See text for details.
      }\label{fig:YORPmap}
\end{figure*}

Figure \ref{fig:YORPmap} 
shows isochrones for $t$ = 0.1, 1, 10, 100, and 1,000\,Myr.
Based on the isochrones, 
tiny NEOs with diameters 
less than 10\,m and ages older than 10\,Myr, 
corresponding to the typical dynamical evolution timescale of the NEOs,
rotates faster than about 10\,s.
However, such fast rotators are not found other than $2020\,\mathrm{HS_7}$.
The observed truncation is not produced by the constant acceleration model.

The densities and surface properties of NEOs depend on their 
sizes \citep{Carry2012}.
However, the density difference of NEOs is a factor of a few at most and does not 
suppress the acceleration of rotation sufficiently.
The thermal inertia also
has little effect on the rotational period
({\v{C}}apek \& Vokrouhlick{\'{y}}\,\yearcite{Capek2004};\, \cite{Golubov2021}).
Therefore, other dynamical mechanisms are required to explain the flat-top distribution.

\subsubsection{Tensile strength}
We discuss the possibility that fast-rotating tiny asteroids are destroyed by the centrifugal force.
The critical rotational period for keeping the shape against the centrifugal force, $P_{\mathrm{cri}}$,
is expressed as follows:
\begin{eqnarray}
    & P_{\mathrm{cri}}=0.42{(C_{\mathrm{shape}})}^{-1}
    \left(\frac{\rho}{2500\,\mathrm{kg\,m^{-3}}}\right)^{1/2} 
    \nonumber \\
    & \times 
    \left(\frac{\kappa}{10^5\,\mathrm{N\,m^{-3/2}}}\right)^{-1/2}
    \left(\frac{D}{1\,\mathrm{m}}\right)^{5/4}\,\,\mathrm{s},
    \label{eq:Pcri}
\end{eqnarray}
where $\rho$ is a bulk density and $\kappa$ is a tensile strength 
coefficient \citep{Holsapple2007, Kwiatkowski2010}.
$C_{\mathrm{shape}}$ is the coefficient indicating the shape of the asteroid defined as 
\begin{equation}
    C_{\mathrm{shape}} = (C_1 C_2)^{1/3}\sqrt
    {\frac{5(3C_{\mathrm{fric}}(1+C_2^2)-\sqrt{3(1-C_2^2+C_2^4)})}{3{C_{\mathrm{fric}}}^2(1+C_2^2)^2-1+C_2^2-C_2^4}},
\end{equation}
where $C_1$ and $C_2$ are the axial ratios of $c/a$ and $b/a$, respectively, 
and $C_\mathrm{fric}$ is a friction coefficient \citep{Holsapple2007}.
We adopt that $C_1$ is 0.7, $C_2$ is 0.7, and $C_\mathrm{fric}$ is 0.31 corresponding to 
a friction angle of $40^{\circ}$.
Then, the shape coefficient $C_\mathrm{shape}$ equals 1.8.

We present two lines indicating $P_{\mathrm{cri}}$ with tensile strength of 
typical stony meteorites ($\kappa = 10^5\,\mathrm{N\,m^{-3/2}}$, \cite{Kwiatkowski2010})
and weak material ($\kappa = 10^3\,\mathrm{N\,m^{-3/2}}$), 
respectively, in panel (b) of figure \ref{fig:YORPmap}.
We use a typical density of S-type asteroids ($\rho=2500\,\mathrm{kg\,m^{-3}}$).
In the case of weak material, 
we can explain the deficiency
of NEOs with $D \leq 10\,\mathrm{m}$ and $P \leq 10\,\mathrm{s}$.
However, the flat-top shape of the distribution is not reproduced 
because $P_\mathrm{cri}$ is proportional to $D^{5/4}$.

\subsubsection{Suppression of YORP by meteoroid impacts}
The YORP acceleration 
can be suppressed by meteoroid impacts onto an asteroid surface
\citep{Farinella1998, Wiegert2015}.
We investigate the evolution of the rotational period taking into account 
possible effects by meteoroid impacts.
We discuss two effects related to meteoroid impacts:
angular momentum transfer and cratering.

\paragraph{Angular momentum transfer}
The absolute angular momentum of an asteroid 
is written as $L = I\omega$, 
where $I$ is a moment of inertia of the asteroid. 
A change of the angular momentum caused by a collision of a meteoroid 
is expressed as
$\delta L = \beta |m \vec{v}_{\mathrm{imp}}\times \vec{R}|$,
where $\beta$ is a dimensionless momentum multiplication factor, 
$m$ is the mass of the meteoroid,
$\vec{v}_{\mathrm{imp}}$ is the impact velocity vector of the meteoroid, and
$\vec{R}$ is the position vector from the center of the asteroid.
Assuming the angle between $\vec{v}_\mathrm{imp}$ and $\vec{R}$ is $90^{\circ}$,
a relative angular momentum change in a single collision is written as
\begin{eqnarray}
    \frac{\delta L}{L} = 
    \frac{\beta m v_{\mathrm{imp}} R}{\frac{2}{5}MR^2\frac{2\pi}{P}}=
    \frac{15 \beta m v_{\mathrm{imp}} P}{16\pi^2\rho R^4},
\end{eqnarray}
where $M$ and $\rho$ are the mass and the bulk density of the asteroid, respectively \citep{Wiegert2015}.
When a collision with a large $\delta L/L$ occurs,
the spin axis of the asteroid can be tilted.
heading to a ceasing of the YORP acceleration. 
Therefore, a timescale of such a critical collision, $\tau_\mathrm{L}$,
corresponds to the duration of the YORP acceleration,

Campbell-Brown \& Braid\,(\yearcite{Campbell-Brown2011})
estimated a flux of meteoroids
from observations of sporadic meteors as
\begin{eqnarray}
N(>m) = 
5\times10^{-11}
\left(\frac{m}{2\times 10^{-6}\,\mathrm{kg}}\right)^{-1}
\,\mathrm{m^{-2}s^{-1}} 
\label{eq:flux_meteor}.
\end{eqnarray}
A typical timescale 
that a meteoroid with the mass larger than $m$ 
collides with the asteroid with the radius of $R$ is written as follows:
\begin{eqnarray}
 & \tau_{\mathrm{L}} = \frac{1}{N(>m)\times \pi R^2} \nonumber \\
& = 1.1
\left(\frac{\delta L}{L}\right)
\left(\frac{\beta}{20}\right)^{-1} 
\left(\frac{v_{\mathrm{imp}}}{3\times10^4\,\mathrm{m\,s^{-1}}}\right)^{-1} 
\nonumber \\
& \times
\left(\frac{\rho}{2500\,\mathrm{kg\,m^{-3}}}\right) 
\left(\frac{D}{1\,\mathrm{m}}\right)^2
\left(\frac{P}{1\,\mathrm{s}}\right)^{-1}\,\mathrm{Myr}.
\end{eqnarray}

We adopt that $\beta$ is 20, $v_{\mathrm{imp}}$ is $3\times10^4\,\mathrm{m\,s^{-1}}$,
and $\rho$ is $2500\,\mathrm{kg\,m^{-3}}$ as typical quantities.
A timescale $\tau_{\mathrm{L}}$ provides the possible fastest rotational 
period accelerated by YORP.
We present three limiting lines for different $\delta L/L$ values in panel (c) of figure \ref{fig:YORPmap}.
The YORP acceleration of smaller asteroids are more suppressed 
by the angular momentum transfer.
However, the flat-top shape of the distribution is not reproduced because 
the reachable periods are proportional to $D$.

\paragraph{Cratering}
When a sufficiently large fraction of the asteroid surface is 
covered with the craters, the continuous YORP acceleration is not an
appropriate assumption since YORP is sensitive to small structures \citep{Statler2009}.
To discuss the cratering effect by meteoroid impacts,
we use the crater scaling law in \citet{Holsapple1993}:
\begin{equation}
    \frac{\rho V_{\mathrm{crater}}}{m} = K_2 \left(\frac{Y}{\rho v_{\mathrm{imp}}^2}\right)^{-3\mu/2}
    \label{eq:crater_scaling},
\end{equation}
where $V_{\mathrm{crater}}$ is the volume of the crater, 
$Y$ is the tensile strength of the target,
and both $K_2$ and slope $\mu$ are constants depending on the taxonomic type of the target.
We refer to the material strength in \citet{Holsapple2020}:
\begin{equation}
    Y = 1.5 \times 10^7 \left(\frac{D}{10^{-1}\,\mathrm{m}}\right)^{1/4} 
    \,\,\mathrm{Pa}.
\end{equation}

The radius of the crater $R_{\mathrm{crater}}$ is written as follows:
\begin{equation}
    R_{\mathrm{crater}} = K_R V_{\mathrm{crater}}^{1/3}\,\,\mathrm{m}
    \label{eq:crater_radius},
\end{equation}
where $K_R$ is a constant which depending on the crater shape.

From equations (\ref{eq:crater_scaling})--(\ref{eq:crater_radius}), 
a surface area of a single crater, $S_\mathrm{crater}$, is written as follows
assuming a bowl-like crater:
\begin{equation}
    S_{\mathrm{crater}} \sim \pi R_{\mathrm{crater}}^2 
    = \pi K_R^2 \left(\frac{K_2}{\rho}\right)^{2/3} 
    \left(\frac{Y}{\rho v_{\mathrm{imp}}^2}\right)^{-\mu}
    m^{2/3}\,\,\mathrm{m^2}.
\end{equation}

From the equation (\ref{eq:flux_meteor}),
the flux density of the meteoroids colliding with the target, $n$, 
is expressed as a function of the mass of the impactor, $m$, 
and the radius of the target, $R$, as follows:
\begin{equation}
    n = 10^{-16}m^{-2}\cdot 4\pi R^{2}
    \,\,\mathrm{kg^{-1}\,s^{-1}}.
\end{equation}

We can estimate the total surface of cratering area by meteoroids per unit time,
$S_{\mathrm{crater}}^{\mathrm{total}}$, as follows:
\begin{eqnarray}
    & S_\mathrm{crater}^\mathrm{total}
= \int^{m1}_{m0} S_{\mathrm{crater}} n dm 
= 12\pi^2\times 10^{-16}
\nonumber \\ 
& \times (m_0^{-1/3}-m_1^{-1/3}) R^2K_R^2 \left(\frac{K_2}{\rho}\right)^{2/3}
\left(\frac{Y}{\rho v_{\mathrm{imp}}^2}\right)^{-\mu}
    \,\,\mathrm{m^{2}\,s^{-1}},
\end{eqnarray}
where $m_0$ and $m_1$ are 
minimum and maximum masses of
the meteoroids, respectively.
We set $m_1 \to \inf$ and $m_0$ as a free parameter.

We assume that no further YORP acceleration works 
once the craters cover certain fraction of the target surface, $\delta S/S$.
The timescale covering $\delta S/S$ of the surface with craters, 
$\tau_\mathrm{crater}$, is
expressed as 
\begin{eqnarray}
\tau_\mathrm{crater} = 
\left(\frac{\delta S}{S}\right) 
\frac{10^{16}}{3\pi K_R^2} 
\left(\frac{\rho}{K_2}\right)^{2/3} 
\left(\frac{Y}{\rho v_{\mathrm{imp}}^2}\right)^{\mu}
m_0^{1/3}
\,\,\mathrm{Myr},
\end{eqnarray}
where $S$ in the entire surface area of the target.

We adopt that $K_2$ is 1 and $\mu$ is 0.55, typical values for S-type asteroids.
Assuming a bowl-like crater, we set $K_R$ to 1.3 \citep{Holsapple1993, Holsapple2020}.
The timescale for S-type asteroids, $\tau_\mathrm{crater, S}$, is given by
\begin{eqnarray}
& \tau_\mathrm{crater, S} = 
0.13
\left(\frac{\rho}{2500\,\mathrm{kg\,m^{-3}}}\right)^{2/3-0.55}  
\left(\frac{v_{\mathrm{imp}}}{3\times10^4\,\mathrm{m\,s^{-1}}}\right)^{-1.1} 
\nonumber \\
& \times 
\left(\frac{D}{1\,\mathrm{m}}
\right)^{0.55/4} 
\left(\frac{m_0}{2\times10^{-6}\,\mathrm{kg}}
\right)^{1/3} 
\left(\frac{\delta S}{S}\right) 
\,\,\mathrm{Myr}.
\end{eqnarray}

We adopt
$\rho$ of $2500\,\mathrm{kg\,m^{-3}}$, 
$v_{\mathrm{imp}}$ of $3\times10^4\,\mathrm{m\,s^{-1}}$.
The possible fastest rotational periods are presented 
over a wide range of $m_0$ values in panel (d) of figure \ref{fig:YORPmap}.
The possible fastest rotational period is approximately proportional to $D$.
Therefore, we cannot explain the truncated distribution
with the suppression of YORP by cratering.

\subsubsection{Tangential YORP effect}
We have considered only the normal YORP (NYORP) disregarding tangential YORP (TYORP).
TYORP depends on the rotational period and thermal properties of the asteroid as with NYORP.
In most cases, TYORP contributes to the acceleration of the rotation unlike NYORP, 
which decelerates the rotation as well
(Golubov \& Kruguly\,\yearcite{Golubov2012};\, \cite{Golubov2014}). 

By taking both NYORP and TYORP into consideration, 
the YORP acceleration is expressed as follows:
\begin{eqnarray}
    & \frac{d\omega}{dt} = 
    8.5 \times 10^{-18}
    \left(\frac{a_{\mathrm{Bennu}}^2 \sqrt{1-e_{\mathrm{Bennu}}^2}}
    {a_{\mathrm{ast}}^2 \sqrt{1-e_{\mathrm{ast}}^2}}\right)
    \nonumber \\
    & \times 
    \left(\frac{D_{\mathrm{Bennu}}}{D}\right)^2 
    (\gamma + (1-\gamma)\eta(\theta))
\,\,\mathrm{rad\,s^{-2}},
    \label{eq:dw_TYORP}
\end{eqnarray}
where $\gamma$ is a fraction of the NYORP contribution to the
total YORP strength, 
$\eta(\theta)$ is an efficiency function of TYORP,
and $\theta$ is a thermal parameter corresponding to a ratio of 
two characteristic scales related to thermal conductivity:
a thermal conductivity length $L_\mathrm{cond}$ and a length of 
heat conductivity wave $L_\mathrm{wave}$
(Golubov \& Kruguly\,\yearcite{Golubov2012}).

The thermal conductivity length is defined as
\begin{equation}
    L_{\mathrm{cond}} = \frac{\lambda}{{((1-A)^3 \Phi^3 \varepsilon \sigma)}^{1/4}}\,\,\mathrm{m},
\end{equation}
where 
$\lambda$ is the heat conductivity of the asteroid,  
$\varepsilon$ and $A$ are
the thermal emissivity and the Bond albedo of the surface, respectively, 
$\Phi$ is the solar energy flux,
and $\sigma$ is Stefan-Boltzmann's constant.
$L_{\mathrm{cond}}$ is a typical scale how far the heat conduction takes place.
The length of heat conductivity wave is defined as
\begin{equation}
    L_{\mathrm{wave}} = \left({\frac{\lambda}{C \rho \omega}}\right)^{1/2}\,\,\mathrm{m},
\end{equation}
where 
$C$ is the heat capacity of the asteroid.
$L_{\mathrm{wave}}$ is a typical scale how far the heat is transferred 
when considering a time variation against a heat source.

Therefore, $\theta$ is written as follows:
\begin{equation}
  \theta(\omega) = \frac{L_{\mathrm{cond}}}{L_{\mathrm{wave}}} 
  = \frac{(C\rho \lambda \omega)^{1/2}}{((1-A)^3\Phi^3\varepsilon\sigma)^{1/4}}.
\end{equation}
The parameter $\theta$ characterizes 
the temperature condition of the surface
and is a function of a rotational period.

\begin{figure*}
  \includegraphics[width=160mm]{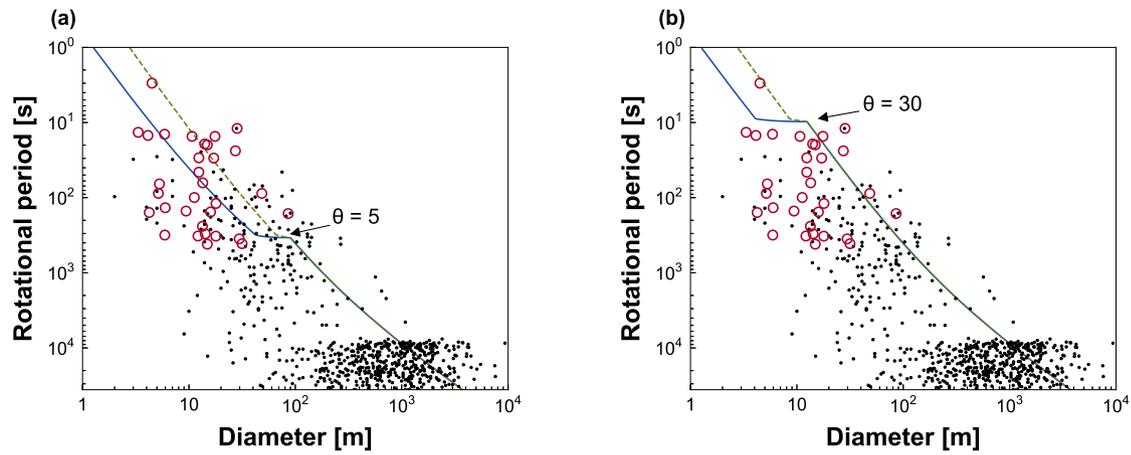}
    \caption{%
      D-P relations of NEOs with isochrones considering the TYORP effect.
      The Tomo-e NEOs and the NEOs in LCDB are presented with
      open and filled circles, respectively.
      (a) Case with $\theta_{\mathrm{max}}$ of 5. (b) Case with $\theta_{\mathrm{max}}$ of 30.
      Isochrones of 10\,My old with $\gamma$ of 0.1 and 0.5 are shown with solid and dashed lines, respectively.
      Arrows indicate positions where $\theta$ reaches $\theta_{\mathrm{max}}$.
      }\label{fig:TYORPmap}
\end{figure*}

Numerical simulations show that the TYORP effect is 
significant for $\theta \sim 1$ (Golubov \& Kruguly\,\yearcite{Golubov2012};\, \cite{Golubov2014}). 
We simplify the efficiency of TYORP
as follows:
\begin{eqnarray}
  \eta(\theta) = 
    1\quad \theta_{\mathrm{min}}<\theta<\theta_{\mathrm{max}}, \\
    0\quad \mathrm{otherwise},
\end{eqnarray}
where $\theta_{\mathrm{min}}$ and $\theta_{\mathrm{max}}$ are
free parameters.
The isochrones considering TYORP 
are shown in panel (a) of figure \ref{fig:TYORPmap}. 
We adopt $\gamma$ equals 0.1 and 0.5 
since TYORP is thought to be as strong as or stronger than NYORP
(Golubov \& Kruguly\,\yearcite{Golubov2012};\, \cite{Golubov2014}). 
Previous studies suggest that parts of tiny NEOs have fine particles 
on the surface \citep{Mommert2014, Fenucci2021}.
We assume that 
the asteroid surface is covered by regolith with 
$\lambda=0.0015\,\mathrm{W\,m^{-1}\,K^{-1}}$,
$C=680\,\mathrm{J\,kg^{-1}\,K^{-1}}$,
and $\rho=1500\,\mathrm{kg\,m^{-3}}$.
We set $\varepsilon=0.7$ and $a=2.0\,\mathrm{au}$
corresponding to typical values for NEOs. 
We adopt that $A=0.084$, which is derived with typical properties of 
moderate albedo asteroids: $p_V$ of 0.2 and a phase integral $q$ of 0.42 \citep{Shevchenko2019}. 
As of November 2021, the change of the rotational periods of 
10 asteroids have been confirmed (\cite{Durech2022}, and references therein). 
The range of $\theta$ among the 10 asteroids is calculated to be 0.26 to 1.3. 
Since all the 10 asteroids are accelerated, not decelerated,
we assume that TYORP is effective for all of them.
Thus, we set $\theta_\mathrm{min}$ and $\theta_\mathrm{max}$ to 0.1 and 5, respectively.
The YORP acceleration considering TYORP successfully leads to flat-top shapes 
around $D\sim100\,\mathrm{m}$ and $P\sim300\,\mathrm{s}$ since the YORP 
acceleration become weaker at $\theta = \theta_{\mathrm{max}}=5$.
However, they do not match the truncated distribution 
around $D\sim10\,\mathrm{m}$ and $P\sim10\,\mathrm{s}$
seen in the D-P relation diagram. 
In the case of $\theta_\mathrm{max}=30$, 
the isochrones become similar to the observed distribution as shown in figure \ref{fig:TYORPmap} (b). 
The larger $\theta$ value means that much more asteroids experience 
TYORP acceleration than theoretical predictions. 
The fact that TYORP acts up to $\theta=30$ 
under assumptions above is rephrased as 
the asteroids are illuminated by stronger radiation
or
thermal parameters such as $C$ and $\lambda$ are smaller.
Most of Tomo-e NEOs have smaller perihelion distances ($r\sim1\,\mathrm{au}$) 
than $2\,\mathrm{au}$ at the time of observations.
Thus, in the case of $C$ or $\lambda$ is an order of magnitude less, 
$\theta$ changes by the factor $(2^2)^{3/4}\times(10)^{1/2}\sim9$.
The observed truncation around 10\,s in rotational period may be produced by the 
TYORP effect.

\section{Conclusions}
A rotational period of an asteroid reflects its dynamical history and physical properties.
We have obtained the light curves of 60 tiny (diameter less than 100\,m) NEOs 
with the wide-field CMOS camera Tomo-e Gozen.
We successfully derived the rotational periods 
and axial ratios of 32 samples owing to the 
video observations at 2\,fps.
We found 13 objects with rotational periods less than 60\,s. 
Compared with literature, 
the distribution of the rotational periods of 32 objects
shows a potential excess in shorter periods.
This result suggests that previous studies missed some population
of fast-rotating asteroids due to long exposure time observations.

We discovered that the distribution of the tiny NEOs 
in the D-P diagram is truncated around a period of 10\,s.
We performed model calculations taking into account the YORP effect.
A NEO smaller than 10\,m is expected to rotate with a period shorter than 
10\,s assuming a constant acceleration by YORP,
which is not consistent the present results.
The truncated distribution is not well explained 
by either the realistic tensile strength of NEOs or the suppression of YORP by meteoroid impacts. 
We found that the tangential YORP effect is a 
possible mechanism to produce the truncated distribution, 
although further observational and theoretical studies as well as high-speed light curve 
observations of NEOs are necessary to reach the conclusion.

\begin{ack}
A special gratitude we give to Mr. Yuto Kojima for his technical 
assistance with this study.
We would like to thank near-Earth asteroid observers around the world.
J. B. would like to express the gratitude to the Iwadare Scholarship Foundation 
and the Public Trust Iwai Hisao Memorial Tokyo Scholarship Fund for the grants.
This work has been supported by the Japan Society for the 
Promotion of Science (JSPS) KAKENHI grants,
21H04491, 20H04617, 18H05223, 18H01272, 18H01261, 18K13599,
17H06363, 16H06341, 16H02158, 26247074, and 25103502.
This work is supported in part by 
the Optical and Near-Infrared Astronomy Inter-University Cooperation Program, 
the Ministry of Education, Culture, Sports, Science and Technology 
(MEXT) of Japan,
JST SPRING, Grant Number JPMJSP2108, 
and the UTEC UTokyo Scholarship.
This work has made use of data from the European Space Agency (ESA) mission
{\it Gaia} (\url{https://www.cosmos.esa.int/gaia}), processed by the {\it Gaia}
Data Processing and Analysis Consortium (DPAC,
\url{https://www.cosmos.esa.int/web/gaia/dpac/consortium}).
Funding for the DPAC has been provided by national institutions, in particular
the institutions participating in the {\it Gaia} Multilateral Agreement.
\end{ack}

\clearpage
\appendix
\section*{Light curves, periodograms, and phased light curves} 
Light curves, Lomb-Scargle periodograms, and phased light curves of the Tomo-e NEOs 
are presented in the figures (\ref{fig:rotPest})--(\ref{fig:rotPest_tumbler}).

\begin{figure*}
  \includegraphics[width=14cm]{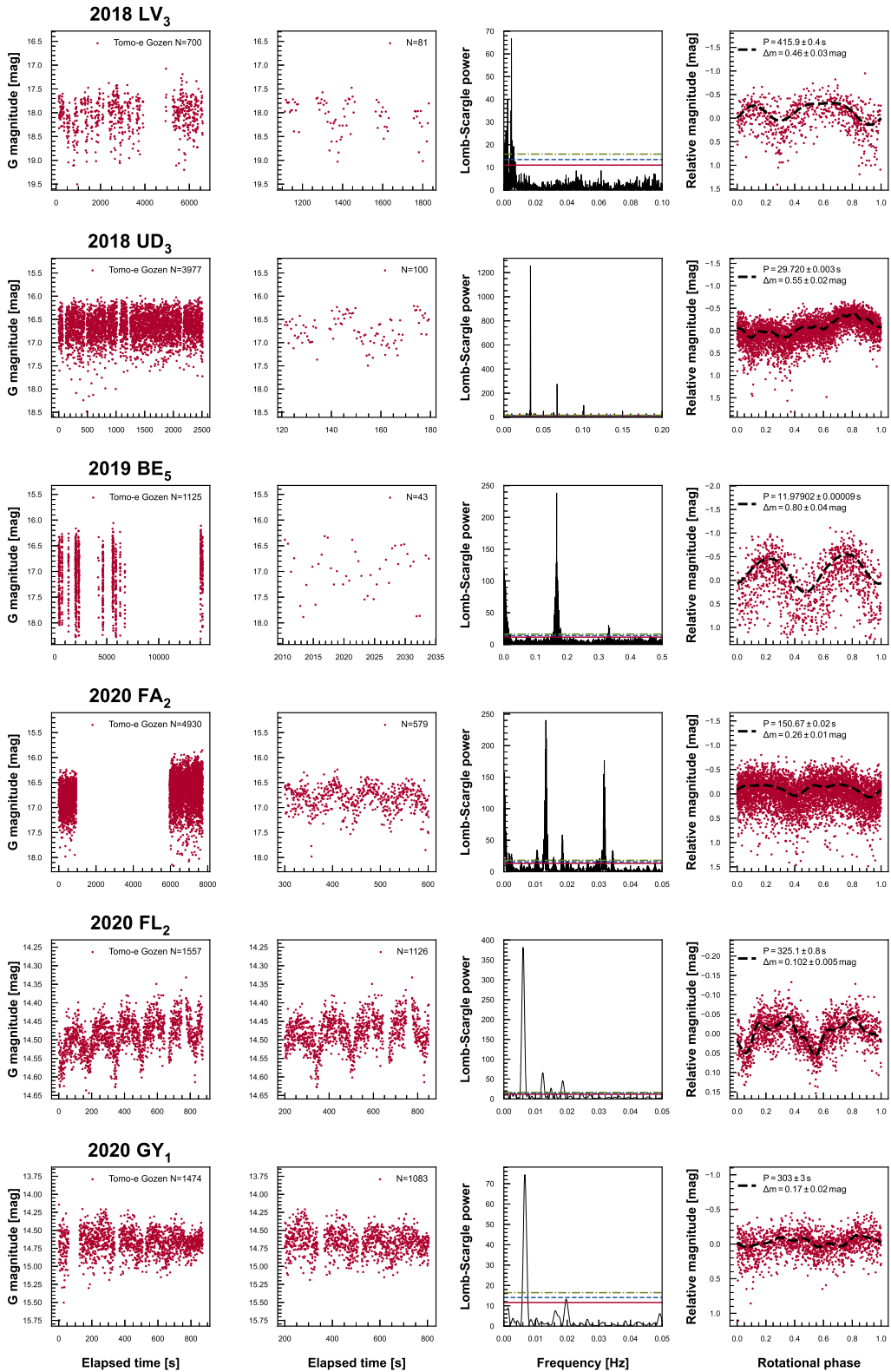}
  \caption{%
    From left to right,
    full light curves from the exposure starting time, 
    partial light curves, Lomb-Scargle periodograms with $n$ of 1,
    and phased light curves of the NEOs whose rotational periods are dirived with
    high reliability.
    Solid, dashed, and dot-dashed horizontal lines in the periodograms
    show 90.0, 99.0, and 99.9 \% confidence levels, respectively. 
    Confidence lines of some NEOs with strong peaks are hard to see due to scale effects.
    Dashed lines in the phased light curves show the model curves.
    Twice the rotational periods are adopted as time ranges of
    the partial light curves.
    }\label{fig:rotPest}
\end{figure*}
\clearpage

\addtocounter{figure}{-1}
\begin{figure*}
  \includegraphics[width=14cm]{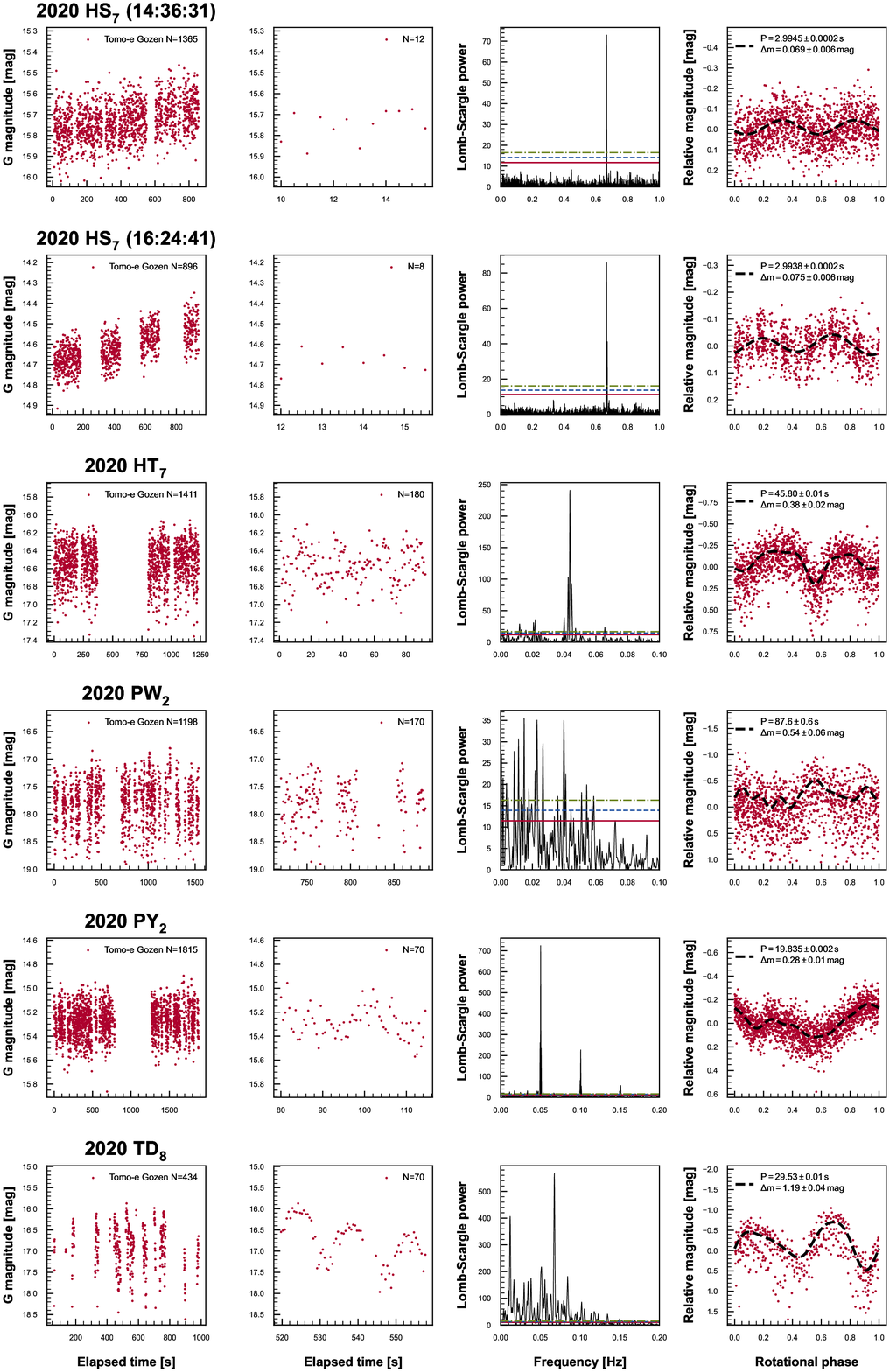}
  \caption{(Continued)}
\end{figure*}
\clearpage

\addtocounter{figure}{-1}
\begin{figure*}
  \includegraphics[width=14cm]{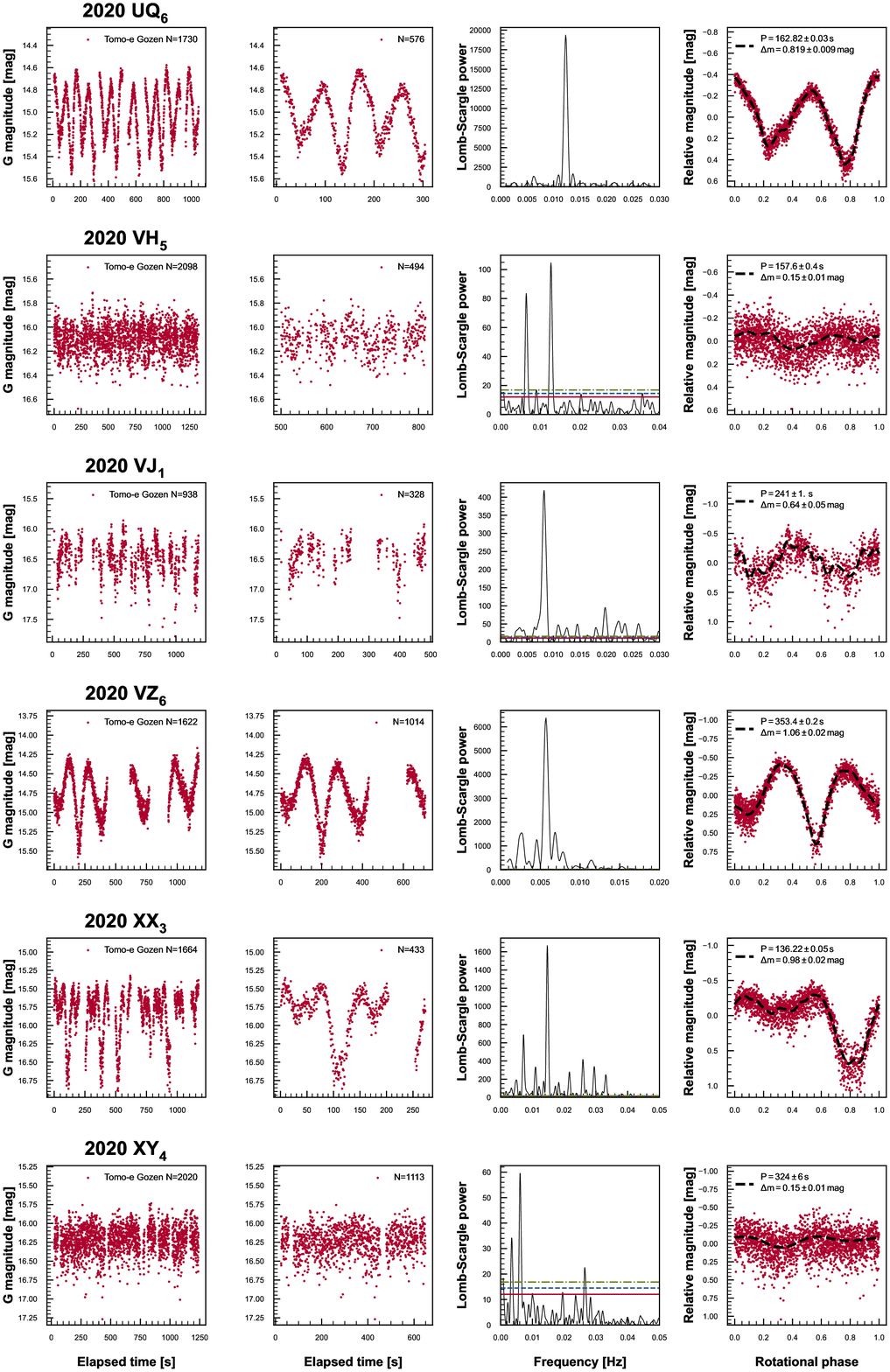}
  \caption{(Continued)}
\end{figure*}
\clearpage

\addtocounter{figure}{-1}
\begin{figure*}
  \includegraphics[width=14cm]{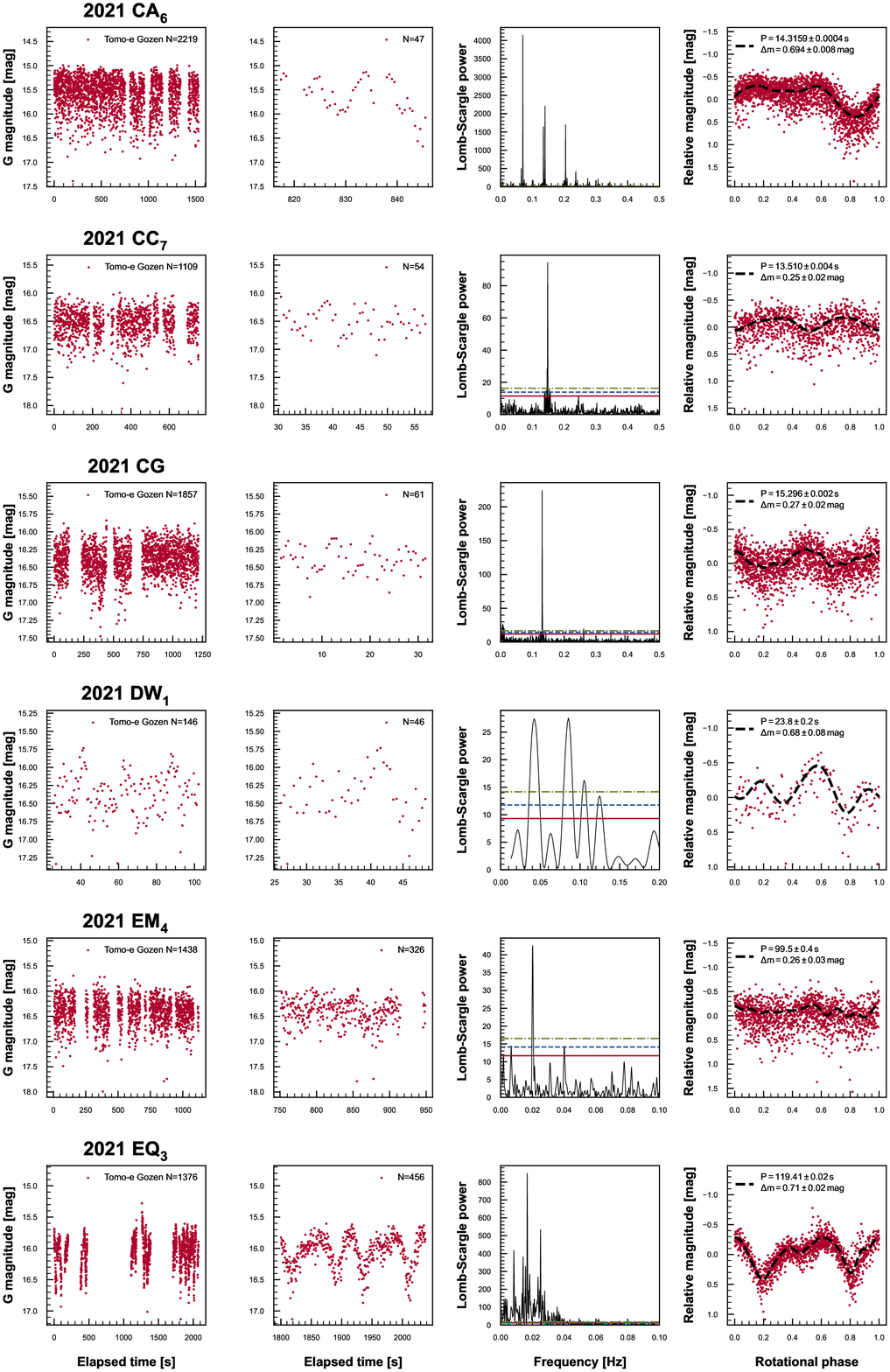}
  \caption{(Continued)}
\end{figure*}
\clearpage

\addtocounter{figure}{-1}
\begin{figure*}
  \includegraphics[width=14cm]{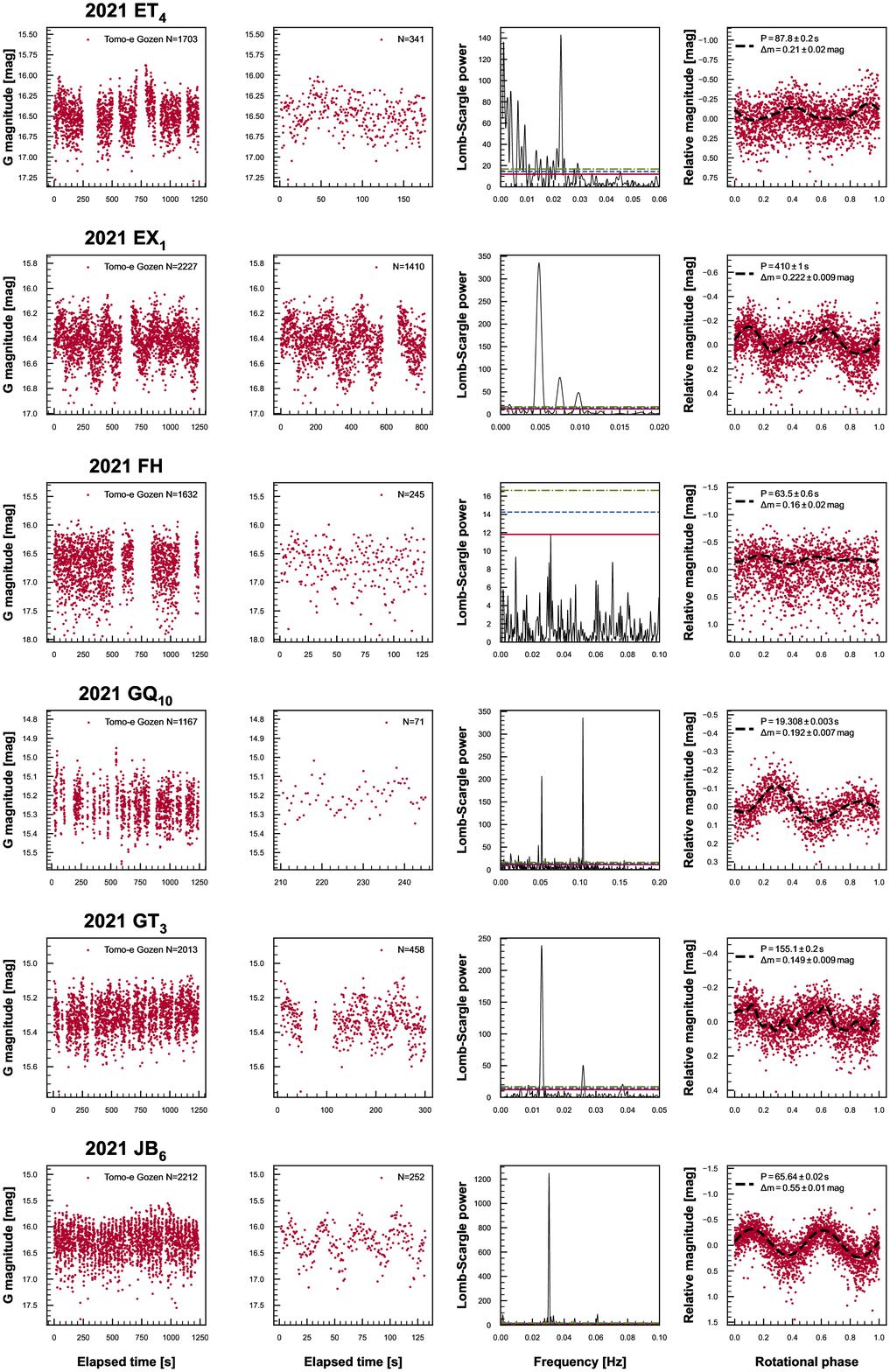}
  \caption{(Continued)}
\end{figure*}
\clearpage

\addtocounter{figure}{-1}
\begin{figure*}
  \includegraphics[width=14cm]{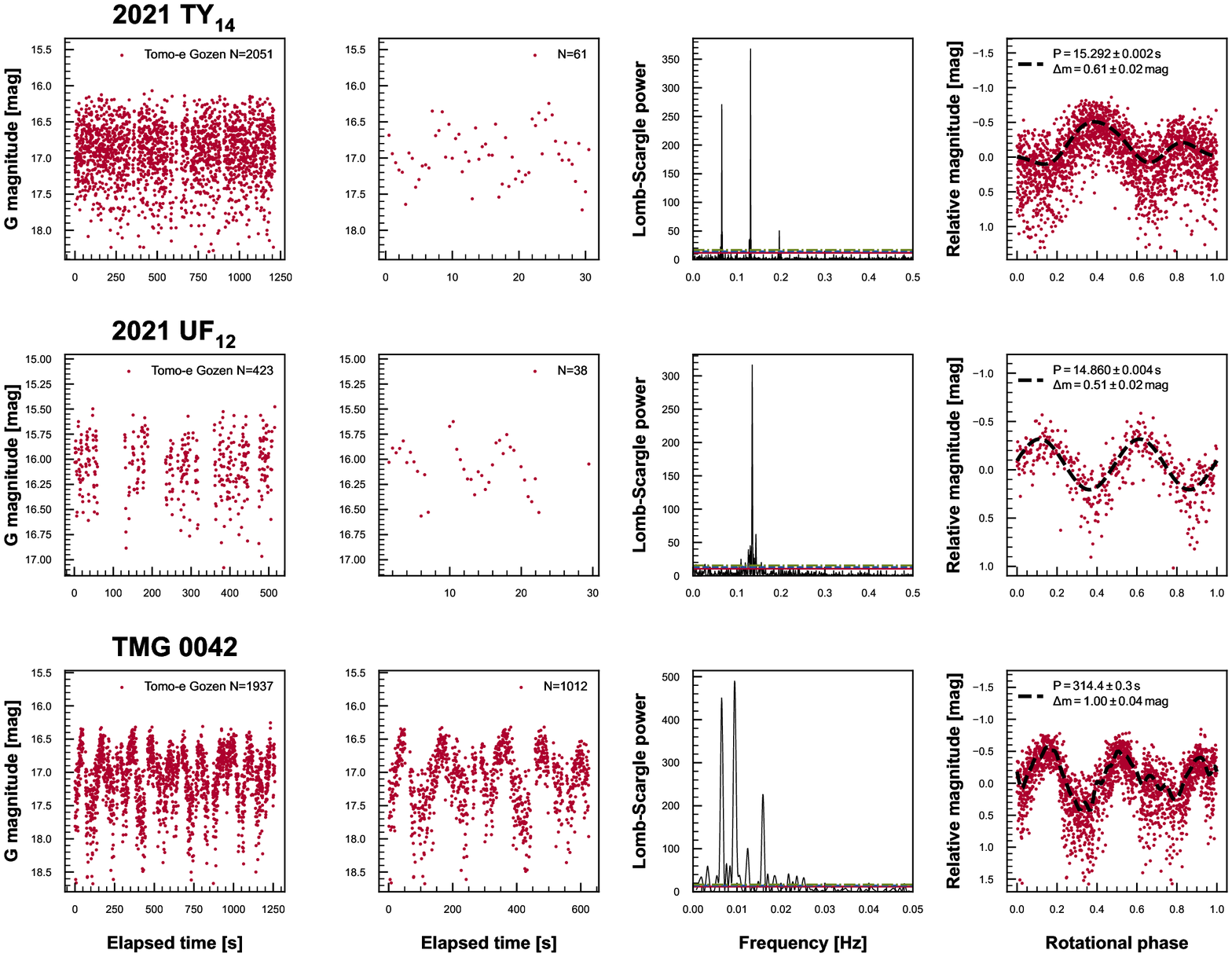}
  \caption{(Continued)}
\end{figure*}
\clearpage

\begin{figure*}
  \includegraphics[width=14cm]{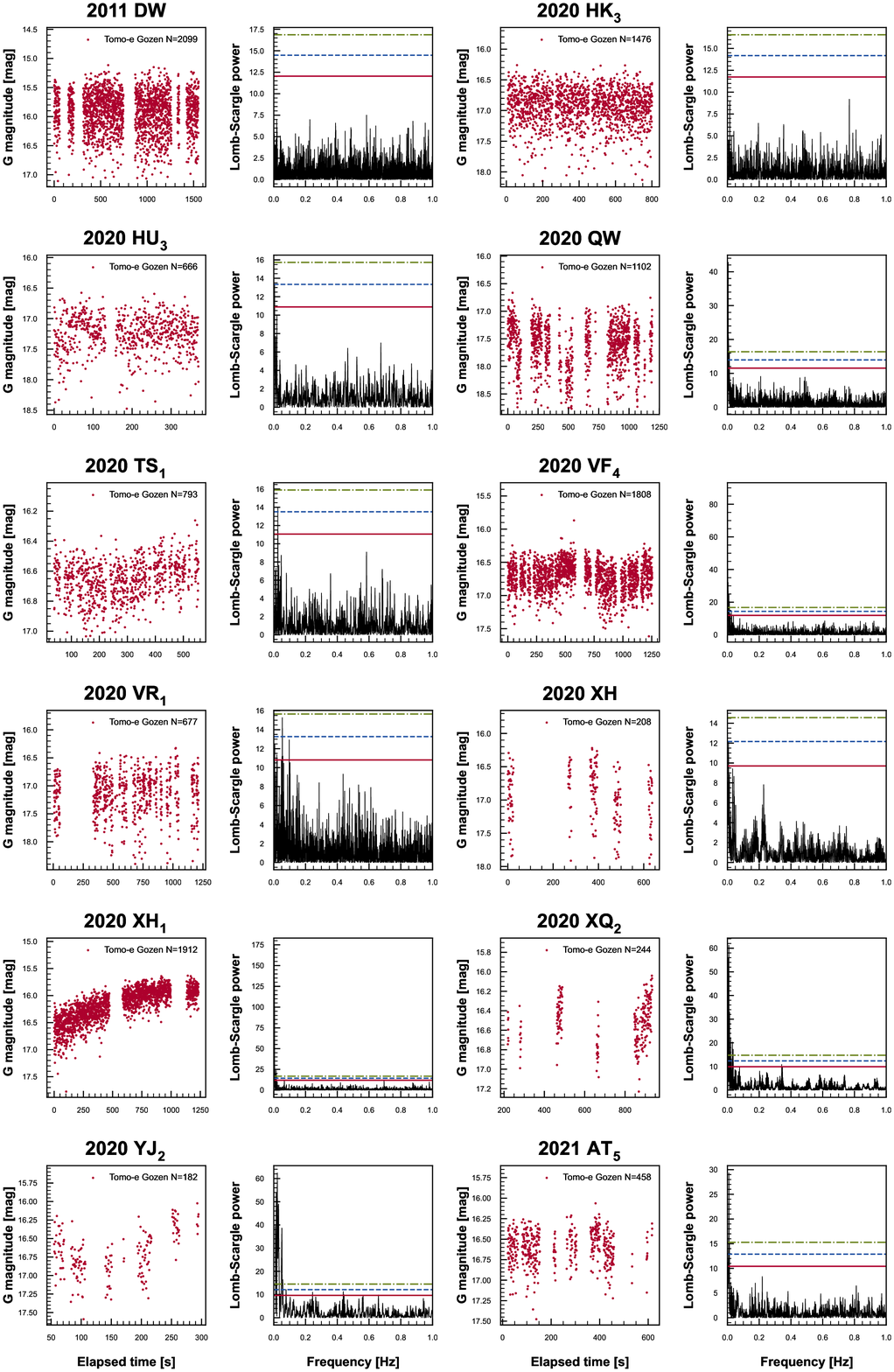}
  \caption{%
    Full light curves and Lomb-Scargle periodograms of 
    the NEOs whose rotational periods are not derived.
    The same as the left 2 columns in figure \ref{fig:rotPest}.
    }
\end{figure*}
\clearpage

\addtocounter{figure}{-1}
\begin{figure*}
  \includegraphics[width=14cm]{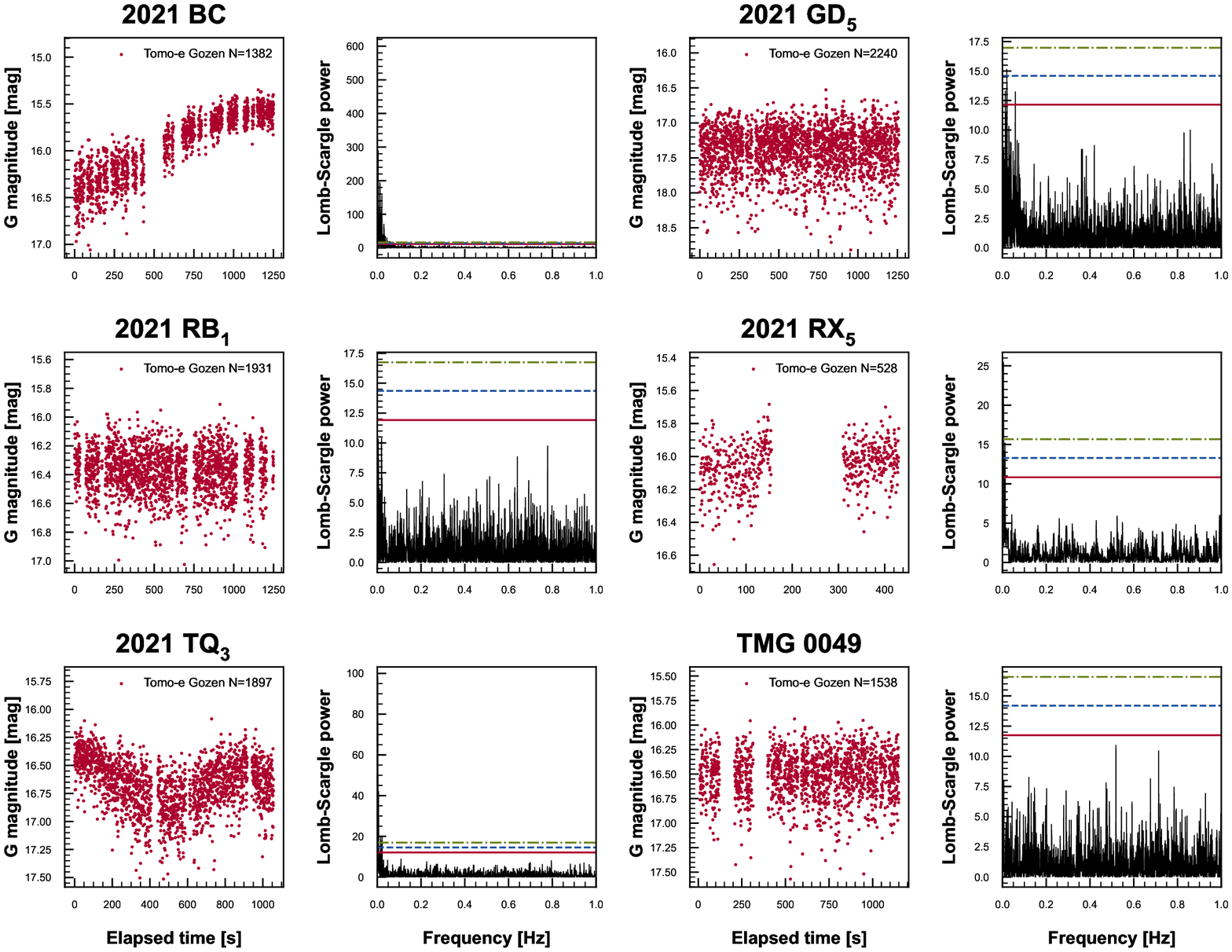}
  \caption{(Continued)}
\end{figure*}
\clearpage

\begin{figure*}
  \includegraphics[width=14cm]{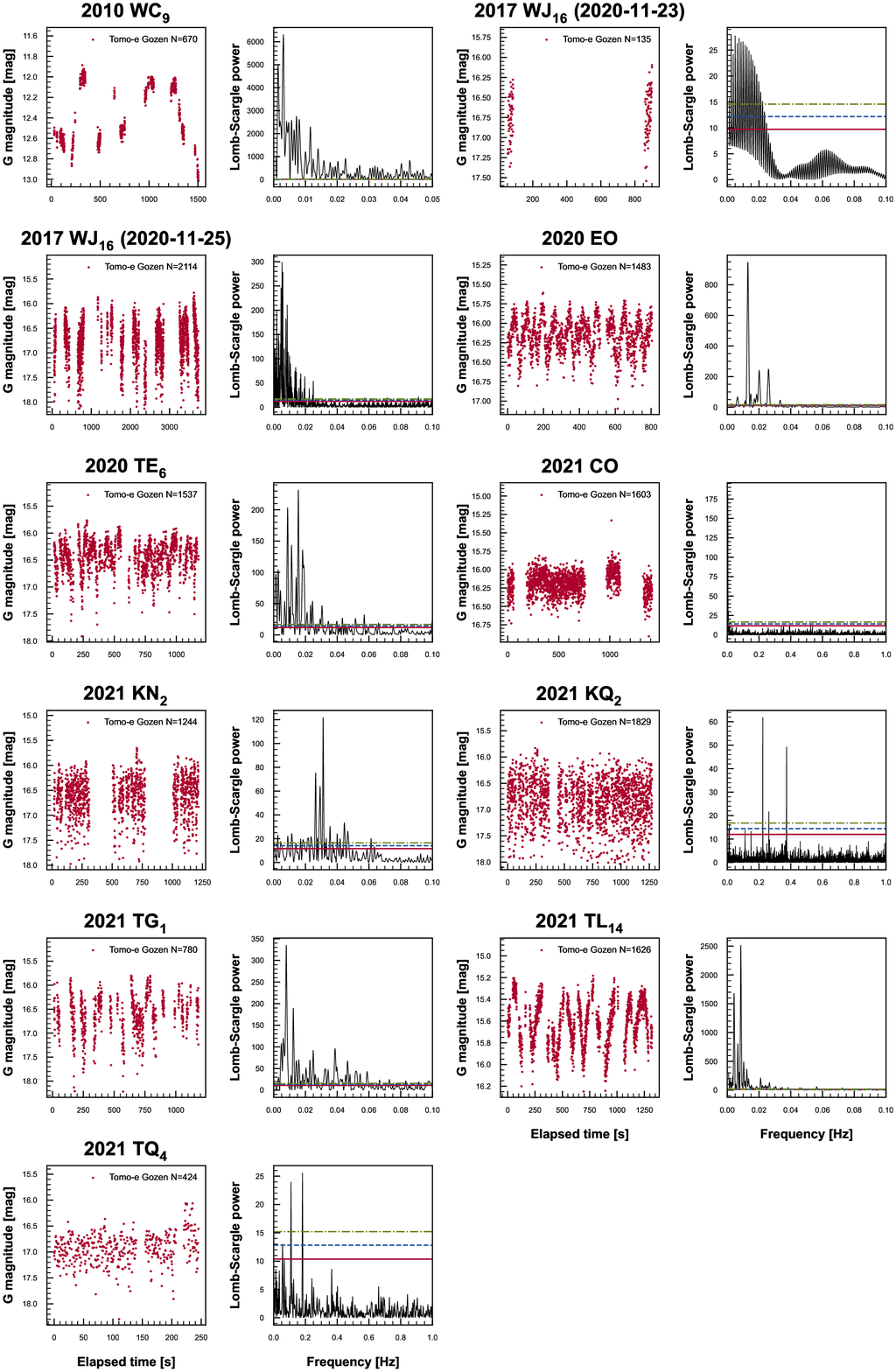}
  \caption{%
    Full light curves and Lomb-Scargle periodograms of 
    the tumbler candidates.
    The same as the left 2 columns in figure \ref{fig:rotPest}.
    }\label{fig:rotPest_tumbler}
\end{figure*}
\clearpage


\begin{thebibliography}{}

\bibitem[{Akaike}(1974)]{Akaike1974}
  {Akaike}, H.\ 1974, IEEE Transactions on Automatic Control, 19, 716

\bibitem[{Barbary} et~al.(2015)]{Barbary2015}
  {Barbary}, K., {Boone}, K., \& {Deil}, C., 2015, {sep: v1.3.0}

\bibitem[{Bertin}, {Arnouts}(1996)]{Bertin1996}
  {Bertin}, E., \& {Arnouts}, S.\ 1996, \aaps, 117, 393

\bibitem[{Birtwhistle}({\noopsort{a}}2021a)]{Birtwhistle2021a}
  {Birtwhistle}, P.\ {\noopsort{a}}2021a, Minor Planet Bulletin, 48, 180

\bibitem[{Birtwhistle}({\noopsort{b}}2021b)]{Birtwhistle2021b}
  {Birtwhistle}, P.\ {\noopsort{b}}2021b, Minor Planet Bulletin, 48, 286

\bibitem[{Birtwhistle}({\noopsort{c}}2021c)]{Birtwhistle2021c}
  {Birtwhistle}, P.\ {\noopsort{c}}2021c, Minor Planet Bulletin, 48, 341

\bibitem[{Bottke} et~al.(2000)]{Bottke2000b}
  {Bottke}, W.~F., {Jedicke}, R., {Morbidelli}, A., {Petit}, J.-M., \&
  {Gladman}, B.\ 2000, Science, 288, 2190

\bibitem[Bottke et~al.(2006)]{Bottke2006}
  Bottke, W.~F., Vokrouhlick{\'{y}}, D., Rubincam, D.~P., \& Nesvorn{\'{y}},
  D.\ 2006, Annual Review of Earth and Planetary Sciences, 34, 157

\bibitem[{Bowell} et~al.(1989)]{Bowell1989}
  {Bowell}, E., {Hapke}, B., {Domingue}, D., {Lumme}, K., {Peltoniemi}, J., \&
  {Harris}, A.~W.\ 1989, in Asteroids II, ed. R.~P., {Binzel}, {et~al.} ,~524

\bibitem[{Breiter} et~al.(2011)]{Breiter2011}
  {Breiter}, S., {Ro{\.z}ek}, A., \& {Vokrouhlick{\'y}}, D.\ 2011, \mnras, 417,
  2478

\bibitem[{Campbell-Brown}, {Braid}(2011)]{Campbell-Brown2011}
  {Campbell-Brown}, M.~D., \& {Braid}, D.\ 2011, in Meteoroids: The Smallest
  Solar System Bodies, ed. W.~J., {Cooke}, {et~al.} ,~304

\bibitem[{\v{C}}apek, Vokrouhlick{\'{y}}(2004)]{Capek2004}
  {\v{C}}apek, D., \& Vokrouhlick{\'{y}}, D.\ 2004, Icarus, 172, 526

\bibitem[{Carry}(2012)]{Carry2012}
  {Carry}, B.\ 2012, \planss, 73, 98

\bibitem[{Chambers} et~al.(2016)]{Chambers2016}
  {Chambers}, K.~C., {et~al.}\ 2016, arXiv:1612.05560

\bibitem[{Drake} et~al.(2009)]{Drake2009}
  {Drake}, A.~J., {et~al.}\ 2009, \apj, 696, 870

\bibitem[{Eugster} et~al.(2006)]{Eugster2006}
  {Eugster}, O., {Herzog}, G.~F., {Marti}, K., \& {Caffee}, M.~W.\ 2006, in
  Meteorites and the Early Solar System II, ed. D.~S., {Lauretta}, \& H.~Y.,
  {McSween},~829

\bibitem[{Farinella} et~al.(1998)]{Farinella1998}
  {Farinella}, P., {Vokrouhlick{\'y}}, D., \& {Hartmann}, W.~K.\ 1998, Icarus,
  132, 378

\bibitem[{Fenucci} et~al.(2021)]{Fenucci2021}
  {Fenucci}, M., {Novakovi{\'c}}, B., {Vokrouhlick{\'y}}, D., \& {Weryk},
  R.~J.\ 2021, \aap, 647, A61

\bibitem[{Fowler}, {Chillemi}(1992)]{Fowler1992}
  {Fowler}, J.~W., \& {Chillemi}, J.~R.\ 1992, Phillips Lab. Tech. Rep., 2049,
  17

\bibitem[{Gaia Collaboration}  et~al.(2018)]{Gaia2018}
  {Gaia Collaboration}, {et~al.}\ 2018, A\&A, 616, A1

\bibitem[{Ginsburg} et~al.(2019)]{Ginsburg2019}
  {Ginsburg}, A., {et~al.}\ 2019, \aj, 157, 98

\bibitem[{Gladman} et~al.(1997)]{Gladman1997}
  {Gladman}, B.~J., {et~al.}\ 1997, Science, 277, 197

\bibitem[{Golubov} et~al.(2014)]{Golubov2014}
  {Golubov}, O., {Scheeres}, D.~J., \& {Krugly}, Y.~N.\ 2014, \apj, 794, 22

\bibitem[{Golubov}, {Krugly}(2012)]{Golubov2012}
  {Golubov}, O., \& {Krugly}, Y.~N.\ 2012, \apjl, 752, L11

\bibitem[{Golubov} et~al.(2021)]{Golubov2021}
  {Golubov}, O., {Unukovych}, V., \& {Scheeres}, D.~J.\ 2021, \aj, 162, 8

\bibitem[{Granvik} et~al.(2018)]{Granvik2018}
  {Granvik}, M., {et~al.}\ 2018, Icarus, 312, 181

\bibitem[{Hatch}, {Wiegert}(2015)]{Hatch2015}
  {Hatch}, P., \& {Wiegert}, P.~A.\ 2015, \planss, 111, 100

\bibitem[Hergenrother  et~al.(2019)]{Hergenrother2019}
  Hergenrother, C.~W., {et~al.}\ 2019, Nature Communications, 10

\bibitem[{Holsapple}(1993)]{Holsapple1993}
  {Holsapple}, K.~A.\ 1993, Annual Review of Earth and Planetary Sciences, 21,
  333

\bibitem[Holsapple(2007)]{Holsapple2007}
  Holsapple, K.~A.\ 2007, Icarus, 187, 500

\bibitem[{Holsapple}(2020)]{Holsapple2020}
  {Holsapple}, K.~A.\ 2020, arXiv:2012.15300

\bibitem[Kadono et~al.(2009)]{Kadono2009}
  Kadono, T., Arakawa, M., Ito, T., \& Ohtsuki, K.\ 2009, Icarus, 200, 694

\bibitem[{Kojima} et~al.(2018)]{Kojima2018}
  {Kojima}, Y., {et~al.}\ 2018, Society of Photo-Optical Instrumentation
  Engineers (SPIE) Conference Series, 10709, 107091T

\bibitem[{Kwiatkowski} et~al.(2021)]{Kwiatkowski2021}
  {Kwiatkowski}, T., {et~al.}\ 2021, \aap, 656, A126

\bibitem[{Kwiatkowski} et~al.(2010)]{Kwiatkowski2010}
  {Kwiatkowski}, T., {Polinska}, M., {Loaring}, N., {Buckley}, D.~A.~H.,
  {O'Donoghue}, D., {Kniazev}, A., \& {Romero Colmenero}, E.\ 2010, \aap, 511,
  A49

\bibitem[{Lomb}(1976)]{Lomb1976}
  {Lomb}, N.~R.\ 1976, \apss, 39, 447

\bibitem[{Michikami} et~al.(2019)]{Michikami2019}
  {Michikami}, T., {et~al.}\ 2019, Icarus, 331, 179

\bibitem[{Michikami} et~al.(2010)]{Michikami2010}
  {Michikami}, T., {Nakamura}, A.~M., \& {Hirata}, N.\ 2010, Icarus, 207, 277

\bibitem[{Mommert} et~al.(2014)]{Mommert2014}
  {Mommert}, M., {et~al.}\ 2014, \apjl, 789, L22

\bibitem[{Ohsawa}(2021)]{Ohsawa2021}
  {Ohsawa}, R.\ 2021, arXiv:2109.09064

\bibitem[{Paolicchi} et~al.(2002)]{Paolicchi2002}
  {Paolicchi}, P., {Burns}, J.~A., \& {Weidenschilling}, S.~J.\ 2002, in
  Asteroids III, ed. J., W.~F., {Bottke}, {et~al.},~517

\bibitem[{Pravec}, {Harris}(2007)]{Pravec2007}
  {Pravec}, P., \& {Harris}, A.~W.\ 2007, Icarus, 190, 250

\bibitem[{Pravec} et~al.(2005)]{Pravec2005}
  {Pravec}, P., {et~al.}\ 2005, Icarus, 173, 108

\bibitem[Pravec, Harris(2000)]{Pravec2000}
  Pravec, P., \& Harris, A.~W.\ 2000, Icarus, 148, 12

\bibitem[Rubincam(2000)]{Rubincam2000}
  Rubincam, D.~P.\ 2000, Icarus, 148, 2

\bibitem[{Sako} et~al.(2018)]{Sako2018}
  {Sako}, S., {et~al.}\ 2018, Society of Photo-Optical Instrumentation
  Engineers (SPIE) Conference Series, 10702, 107020J

\bibitem[{Scargle}(1982)]{Scargle1982}
  {Scargle}, J.~D.\ 1982, \apj, 263, 835

\bibitem[{Shevchenko} et~al.(2019)]{Shevchenko2019}
  {Shevchenko}, V.~G., {et~al.}\ 2019, \aap, 626, A87

\bibitem[{Statler}(2009)]{Statler2009}
  {Statler}, T.~S.\ 2009, Icarus, 202, 502

\bibitem[{Thirouin} et~al.(2016)]{Thirouin2016}
  {Thirouin}, A., {et~al.}\ 2016, \aj, 152, 163

\bibitem[{Thirouin} et~al.(2018)]{Thirouin2018}
  {Thirouin}, A., {et~al.}\ 2018, \apjs, 239, 4

\bibitem[{Tonry} et~al.(2018)]{Tonry2018}
  {Tonry}, J.~L., {et~al.}\ 2018, \pasp, 130, 064505

\bibitem[{VanderPlas}(2018)]{VanderPlas2018}
  {VanderPlas}, J.~T.\ 2018, \apjs, 236, 16

\bibitem[{{\v{D}}urech} et~al.(2022)]{Durech2022}
  {{\v{D}}urech}, J., {et~al.}\ 2022, \aap, 657, A5

\bibitem[{Vokrouhlick{\'y}} et~al.(2007)]{Vokrouhlicky2007}
  {Vokrouhlick{\'y}}, D., {Breiter}, S., {Nesvorn{\'y}}, D., \& {Bottke},
  W.~F.\ 2007, Icarus, 191, 636

\bibitem[Vokrouhlick{\'{y}} et~al.(2000)]{Vokrouhlicky2000}
  Vokrouhlick{\'{y}}, D., Milani, A., \& Chesley, S.~R.\ 2000, Icarus, 148, 118

\bibitem[{Vokrouhlick{\'y}}, {{\v{C}}apek}(2002)]{Vokrouhlicky2002}
  {Vokrouhlick{\'y}}, D., \& {{\v{C}}apek}, D.\ 2002, Icarus, 159, 449

\bibitem[{Vokrouhlick{\'y}} et~al.(2004)]{Vokrouhlicky2004}
  {Vokrouhlick{\'y}}, D., {{\v{C}}apek}, D., {Kaasalainen}, M., \& {Ostro},
  S.~J.\ 2004, \aap, 414, L21

\bibitem[{Vokrouhlick{\'y}}(1998)]{Vokrouhlicky1998}
  {Vokrouhlick{\'y}}, D.\ 1998, \aap, 335, 1093

\bibitem[Warner et~al.(2009)]{Warner2009}
  Warner, B.~D., Harris, A.~W., \& Pravec, P.\ 2009, Icarus, 202, 134

\bibitem[Wiegert(2015)]{Wiegert2015}
  Wiegert, P.~A.\ 2015, Icarus, 252, 22

\bibitem[{Yeomans} et~al.(2000)]{Yeomans2000}
  {Yeomans}, D.~K., {et~al.}\ 2000, Science, 289, 2085

\bibitem[{Zappala} et~al.(1990)]{Zappala1990}
  {Zappala}, V., {Cellino}, A., {Barucci}, A.~M., {Fulchignoni}, M., \&
  {Lupishko}, D.~F.\ 1990, \aap, 231, 548

\bibitem[{Zhai} et~al.(2014)]{Zhai2014}
  {Zhai}, C., {et~al.}\ 2014, \apj, 792, 60

\end{thebibliography}
\bibliographystyle{pasjbib}

\newcommand{\noopsort}[1]{} \newcommand{\singleletter}[1]{#1}

\end{document}